\documentclass[fleqn,usenatbib]{mnras}

\usepackage[T1]{fontenc}

\DeclareRobustCommand{\VAN}[3]{#2}
\let\VANthebibliography\thebibliography
\def\thebibliography{\DeclareRobustCommand{\VAN}[3]{##3}\VANthebibliography}

\usepackage[british]{babel}             
\usepackage{newtxtext}               
\usepackage[slantedGreek]{newtxmath}    
\let\la=\lesssim 
\usepackage{ae,aecompl}
\usepackage{graphicx}
\usepackage{graphics}
\usepackage{adjustbox}  
\usepackage{caption}
\usepackage{xcolor}
\usepackage{xspace} 
\usepackage{gensymb}
\usepackage{hyperref}
\usepackage{nth}
\usepackage{tikz}
\usepackage{pifont}

\graphicspath{./{Figs/}}
\DeclareGraphicsExtensions{.pdf}

\newcommand{\Hii}{H{\sc ii}\ } 
\newcommand{\Te}{\ensuremath{T_\text{e}}\xspace}
\newcommand{\Ne}{\ensuremath{N_\text{e}}\xspace}

\newcommand{\mstar}{M$_{\star}$}

\definecolor{ForestGreen}{HTML}{2e8b21}
\definecolor{Teal}{RGB}{0, 165, 164}

\newcommand{\oiii}{[\ion{O}{iii}]$\lambda5007$}
\newcommand{\oii}{[\ion{O}{ii}]$\lambda3727,29$}

\newcommand{\niibpt}{[\ion{N}{ii}]-BPT}

\def\arcsec{\hbox{$^{\prime\prime}$}}

\newcommand\orcid[1]{\href{http://orcid.org/#1}{\adjustbox{trim={-.15\width} {0\height} {-.15\width} {0\height},clip}{\includegraphics[height=10pt]{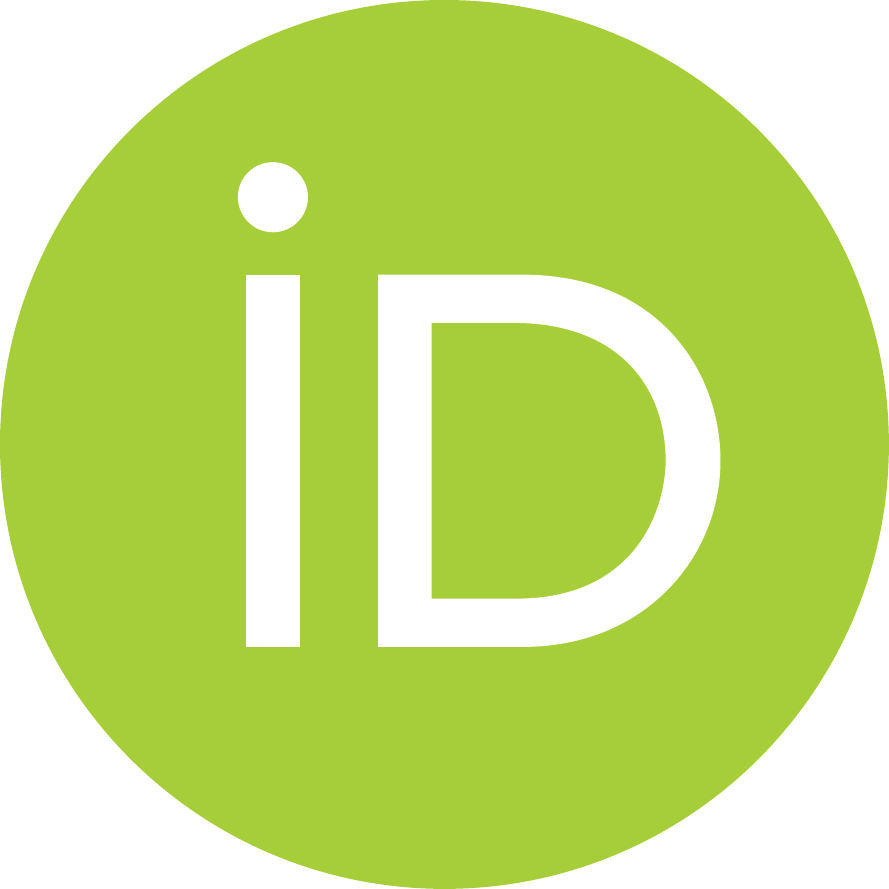}}}}

\title[Direct metallicities with \textit{JWST} at $z\sim8$]{The chemical enrichment in the early Universe as probed by \textit{JWST} via direct metallicity measurements at $z\sim8$}
\author[M.~Curti et al.]{Mirko
 Curti\orcid{0000-0002-2678-2560},$^{1,2}$\thanks{E-mail: mc2041@cam.ac.uk}
 Francesco D'Eugenio\orcid{0000-0003-2388-8172},$^{1,2}$ 
 Stefano Carniani\orcid{0000-0002-6719-380X},$^{3}$ 
 Roberto Maiolino\orcid{0000-0002-4985-3819},$^{1,2,4}$ 
\newauthor
 Lester Sandles\orcid{0000-0001-9276-7062},$^{1,2}$ 
 Joris Witstok\orcid{0000-0002-7595-121X},$^{1,2}$ 
 William M.~Baker\orcid{0000-0003-0215-1104},$^{1,2}$
 Jake S.~Bennett\orcid{0000-0002-8573-2993},$^{1,5}$ 
  \newauthor
Joanna M.~Piotrowska\orcid{0000-0003-1661-2338},$^{1,2}$ 
 Sandro Tacchella\orcid{0000-0002-8224-4505},$^{1,2}$ 
  Stephane Charlot\orcid{0000-0003-3458-2275},$^{6}$
 Kimihiko Nakajima,$^{7}$ 
   \newauthor
  Gabriel Maheson,$^{1,2}$
Filippo Mannucci\orcid{0000-0002-4803-2381},$^{8}$ 
 Amirnezam Amiri\orcid{0000-0002-8553-1964},$^{9,8}$
  Santiago Arribas\orcid{0000-0001-7997-1640},$^{10}$
 \newauthor 
  Francesco Belfiore,$^{8}$
  Nina R.~Bonaventura\orcid{0000-0001-8470-7094},$^{11}$
 Andrew J.~Bunker,$^{12}$ 
 Jacopo Chevallard,$^{12}$ 
  \newauthor 
 Giovanni Cresci,$^{8}$
 Emma Curtis-Lake,$^{13}$
Connor Hayden-Pawson\orcid{0000-0001-7964-1027}$^{1,2}$, 
 Gareth C.~Jones\orcid{0000-0002-0267-9024},$^{12}$ 
  \newauthor 
 Nimisha Kumari,$^{14}$
 Isaac Laseter,$^{15}$
 Tobias J.~Looser\orcid{0000-0002-3642-2446},$^{1,2}$
Alessandro Marconi\orcid{0000-0002-9889-4238},$^{9,8}$ 
   \newauthor 
Michael V.~Maseda\orcid{0000-0003-0695-4414},$^{15}$ 
 Jan Scholtz,$^{1,2}$
  Renske Smit\orcid{0000-0001-8034-7802},$^{16}$
  Hannah \"Ubler\orcid{0000-0003-4891-0794},$^{1,2}$ 
and Imaan E.~B.~Wallace\orcid{0000-0002-0695-8485},$^{12}$ 
\\
\\
\emph{\normalsize Affiliations are listed at the end of the paper}
}

\date{Accepted 2022 September 22. Received 2022 September 22; in original form 2022 July 26
}

\pubyear{2022}

\begin{document}
\label{firstpage}
\pagerange{\pageref{firstpage}--\pageref{lastpage}}
\maketitle

\begin{abstract}
We analyse the chemical properties of three z$\sim$~8 galaxies behind the galaxy cluster SMACS\,J0723.3–7327, observed as part of the Early Release Observations programme of the \textit{James Webb Space Telescope} (\textit{JWST}). 
Exploiting [\ion{O}{iii}]$\lambda4363$ auroral line detections in NIRSpec spectra, we robustly apply the direct \Te method for the very first time at such high redshift, measuring metallicities ranging from extremely metal poor (12+log(O/H)$\approx$~7) to about one-third solar. 
We also discuss the excitation properties of these sources, and compare them with local strong-line metallicity calibrations. 
We find that none of the considered diagnostics match simultaneously the observed relations between metallicity and strong-line ratios for the three sources, implying that a proper re-assessment of the calibrations may be needed at these redshifts. 
On the mass-metallicity plane, the two galaxies at $z\sim7.6$ ($\rm log(M_*/M_{\odot}) = 8.1, 8.7$) have metallicities that are consistent with the extrapolation of the mass-metallicity relation at z$\sim$2-3, while the least massive galaxy at $z\sim8.5$ ($\rm log(M_*/M_{\odot}) = 7.8$) shows instead a significantly lower metallicity .
The three galaxies show different level of offset relative to the Fundamental Metallicity Relation, with two of them (at z$\sim$~7.6) being marginally consistent, while the z$\sim$~8.5 source deviating significantly, being probably far from the smooth equilibrium between gas flows, star formation and metal enrichment in place at later epochs.
\end{abstract}




\section{Introduction}

The study of the abundances of heavy elements (the `metallicity') in the interstellar medium (ISM) of galaxies provides precious
insights on the physical processes responsible for their formation,
and on how the relative importance of such processes has changed across cosmic time \citep[e.g.][ see also the review by \citealt{maiolino_re_2019}]{ma_origin_2016, dave_mufasa_2017, Torrey2019, Langan2020}.

In the local Universe, the metallicity of the gas-phase is observed to correlate tightly with some of the primary galactic properties. Specifically, the metallicity depends on the stellar mass of the galaxy (mass-metallicity relation, MZR, \citealt{tremonti_origin_2004}) and has a secondary (inverse) dependence on the star formation rate (SFR, \citealt{ellison_clues_2008}). This metallicity-$\rm M_\star$-SFR relation has been called the Fundamental Metallicity Relation \citep[FMR,][]{mannucci_fundamental_2010,Mannucci2011,curti_massmetallicity_2020}, a signature of the smooth, long-lasting equilibrium between gas flows and secular evolution \citep[e.g.][]{bouche_cold_accretion_2010,lilly_gas_2013}.

In the past fifteen years, major efforts have been dedicated to explore the metallicity of high-redshift galaxies through massive spectroscopic surveys in the near-IR
\citep[e.g.][among many others]{erb_high_2016,maiolino_amaze_2008,mannucci_lsd_2009,zahid_mass-metallicity_2011,steidel_strong_2014,zahid_universal_2014,strom_nebular_2017, curti_klever_2020,sanders_mosdef_mzr_2021}. 
These works have established that the mass-metallicity relation is already in place up to $z\sim3.3$, and that it evolves with redshift, in the sense that higher redshift galaxies appear metal deficient compared to galaxies of similar stellar mass at later epochs. 
However, when taking into account the secondary dependence of the MZR on the SFR, as described by the FMR framework, any evolution is apparently canceled out at least up to $z\sim3.3$
\citep[e.g.][]{mannucci_fundamental_2010,Cresci2019, sanders_mosdef_mzr_2021}. 
This finding implies that galaxies up to $z\sim3.3$ follow, on average, the same smooth evolution as local galaxies, and that the evolution of the MZR is a consequence of $z\sim2-3$ galaxies having higher SFR (living at the epoch of the peak of the cosmic star formation rate density, \citealt{madau_cosmic_2014}), therefore populating preferentially the low-metallicity region of the same, non-evolving FMR. Together with the existence of the main sequence of star formation \citep[e.g.][and references therein]{noeske_star_2007, sandles_2022, Popesso2022} this is an indication that secular rather than stochastic processes dominate the evolution of galaxies at these epochs.

Pushing the exploration of galaxy metallicities to redshifts higher than $\sim 3.3$ has proven very challenging so far due to the intrinsic technical limitations of current astronomical facilities, as the primary (optical) nebular diagnostics required to measure the gas-phase metallicity from galaxy spectra are redshifted outside the wavebands observable from the ground \citep[but see][]{Troncoso2014,2017ApJ...846L..30S,2021MNRAS.508.1686W}. 

An additional problem is that metallicity determinations generally rely on adopting locally calibrated strong-line diagnostics, as temperature-sensitive auroral lines are usually too faint to be detected, preventing the use of the more robust and `direct' electron-temperature (\Te) method.
However, it is not clear whether the strong-line calibrations derived in the local universe are valid for distant galaxies, which have drastically different properties \citep[][]{maiolino_re_2019}.
To date, auroral lines have been observed only in a small number of intermediate-redshift (z$\sim$1--3) galaxies, and constitute mostly serendipitous, low-significance detections \citep[e.g.][]{christensen_gravitationally_2012,patricio_testing_2018, 2020MNRAS.491.1427S} 

This observational landscape is set to be revolutionised by the advent of the \textit{James Webb Space Telescope} (\textit{JWST}) and its near-IR spectrograph NIRSpec
\citep[][]{Jakobsen2022,Ferruit2022}, which has opened the capability of obtaining multi-object spectroscopic observations in the near-IR from space (up to $5.3 \mu$m), combining high multiplexing with a sensitivity much higher than any current and past facility.

The first \textit{JWST}/NIRSpec spectra have been recently released within the context of the Early Release Observations, and were obtained by targeting galaxies lensed by the cluster SMACS\,J0723.3–7327
\citep{Ebeling2001,Ebeling2007,Ebeling2010,Mann2012,Ebeling2013,Repp2016,Repp2018}, providing clear detections of nebular lines in galaxies out to $z=$8.5, which can be used to characterise the chemical enrichment of the ISM at the Epoch of Reionisation (EoR), and beyond.
Most remarkably, some of these spectra present clear detections of the $[\mathrm{OIII}]\lambda4363$ auroral line, which can be used to robustly measure the gas metallicity with the \Te\ method for the very first time at such high redshifts. 

In this paper, we extract the direct metallicity of such galaxies at $z\sim8$, which have individual detections of
auroral lines, and test whether they follow the same strong-line metallicity calibration as local galaxies. In addition, we investigate whether these galaxies, probing the EoR,
follow the same metallicity scaling relations as lower-redshift galaxies or exhibit evidence for evolution.

This paper is organised as follows. 
In Section~\ref{sec:data} we describe observations, data processing and data analysis. In particular, we discuss the different steps we have implemented to properly process and calibrate the data by using publicly available information.
We also describe our emission lines fitting procedure, and how SED fitting was performed on NIRCAM photometry to derive different physical properties of these galaxies.
In Section~\ref{sec:line_ratios}, we discuss the emission-line properties of our targets, in terms of Balmer decrements and dust attenuation, excitation-diagnostic diagrams, and metallicity properties as derived with the \Te method. We also compare the observed line ratios and metallicity with some of the most widely adopted strong-line abundance diagnostics calibrated on local galaxies. 
In Section~\ref{section:Evolution_of_the_metallicity_scaling_relations}, we discuss the observed metallicity properties of these targets in the framework of the most relevant metallicity scaling relations, i.e. the mass-metallicity and the fundamental metallicity relation. 
Finally, in Section~\ref{sec:conclusions} we report our conclusions.

Throughout this work, we assume a standard lambda cold dark matter cosmology with H$_{0}$=70 km s$^{-1}$, $\upOmega_{m}$=0.3, and $\upOmega_{\Lambda}$=0.7.

\section{Observations, data processing and data analysis}
\label{sec:data}

\subsection{NIRSpec spectroscopy}

\subsubsection{NIRSpec observations}
We use publicly available data from the Early Release Observations of the cluster SMACS\,J0723.3–7327 (Program ID: 2736, \citealt{pontoppidan_ERO_2022}).
The NIRSpec observations consist of two pointings with the same Multi Shutter Array (MSA) configuration but different acquisition stars (and different filters for the acquisition); in the following, we refer to these two observations as Obs~7 and Obs~8.

NIRSpec observations were carried out by using the disperser-filter combinations G235M/F170LP and G395M/F290LP, which cover the wavelength range between 1.66$\mu$m and 5.16$\mu$m and provide spectra with a spectral resolution of $R\sim1000$.
For each observation, three nodding positions of 20 groups and two integrations each were performed for each grating setup. The total exposure time of the two individual pointings is the same and corresponds to 8,840 seconds for each grating.
While shutters were opened on 35 targets, in this paper we only focus on the three targets with the highest redshift: ID 4590 (z$=$8.4953), ID 6355 (z$=$7.6643) and ID 10612 (z$=$7.6592). These redshifts are based on emission-line velocities measured on the \textit{JWST} spectra (see Sect.~\ref{sec:fitting}).

\subsubsection{NIRSpec data reduction}
\label{sec:data_reduction}
 We have retrieved the level 2 data (i.e. count rate maps) from the MAST archive, but we have then reprocessed the data through the GTO pipeline (NIRSpec/GTO collaboration, in prep.). 
As most of the processing steps use the same algorithms that the pipeline used to generate the MAST archive products (see Fig.~11 and section 4.3 of \citealt{Ferruit2022}), 
the resulting 2D spectra are not very different from those provided by the standard MAST pipeline; however, we perform our own optimised extraction aperture and bad/cosmic pixel flagging and masking. 
Most importantly, to mitigate the fact that many of the calibration reference files used by the pipeline may still correspond to ground data or be model based\footnote{\hyperlink{https://jwst-docs.stsci.edu/jwst-calibration-pipeline-caveats}{https://jwst-docs.stsci.edu/jwst-calibration-pipeline-caveats}}, we have generated a correction to the response function by analysing the observation of the spectrophotometric calibration star 2MASS J18083474+6927286, observed during commissioning, and publicly available through programme \textit{JWST} 1128. We note that this star has been subsequently removed from the list of primary calibrators for \textit{JWST}, on account of its variability \citep[$\sigma$ = 0.41~per\ cent,][]{gordon_flux_cal_2022}; however, for our purposes, this adds a negligible amount of uncertainty.
Specifically, the resulting spectrum of the star processed in the same way as the scientific targets was compared with its intrinsic spectrum from STScI \citep[astronomical catalogue CALSPEC]{Bohlin14,Bohlin20} to derive a more accurate flux calibration.

Given that our sources are marginally resolved, we consider both the assumption of point-like source and extended source to correct for path-losses. However, as the sources are fairly compact (especially compared to the point spread function, PSF, at long wavelengths), we use the point-source assumption for our fiducial analysis.
We note that at long wavelengths (F290LP filter, with PSF comparable or larger than the shutter size), which are of greatest interest in this paper, the relative flux correction does not strongly depend on which of the two cases is adopted, while providing an offset of $\sim2$ on the 
absolute calibration; yet, we verified that adopting the extended-source corrections does not significantly affect the main results of the paper.
Background subtraction was performed through the standard technique of subtracting the average of two nodding positions from the other position.

We also inspect the exposures of the individual nods in order to identify artefacts that might have escaped automated flagging and other potential issues. We note that the auroral lines are detected in the individual exposures, hence confirming their detection at high confidence level, with the exception of galaxy ID 4590 (z$=$8.5), for which no emission lines are detected in one of the nods of Obs~7. By inspecting the 2D images before background subtraction, it was found that, although not identified as faulty, shutter [3,27,167], on which source ID 4590 was nodded, did not open in Obs~7. 
As a consequence, for source ID 4590, we use Obs~8 and only the two noddings of Obs~7 for which the source is in open shutters.


Stacking was performed with the GTO pipeline, taking into account the variances and quality arrays of both observations.
%
The resulting spectra are shown in Fig.\ref{fig:fit}.  Note that these spectra are in F$_{\lambda}$ (while the spectra provided by MAST archive are in F$_{\nu}$).
These spectra can be retrieved from a publicly accessible repository\footnote{Available at \url{https://doi.org/10.5281/zenodo.6940561}.} 

\subsubsection{Spectral fitting}
\label{sec:fitting}
In our analysis, we use only data from the G395M/F290LP disperser/filter combination, because it covers the entire rest-frame wavelength range relevant to this work. We measure line fluxes using {\sc ppxf}, the penalised pixel fitting algorithm of \cite{cappellari_improving_2017}. We rebin the spectra to a regular grid in linear velocity space, using the original velocity sampling as uniform pixel size (102~$\mathrm{km\,s^{-1}}$)\footnote{We verified that e.g. the H$\updelta$/H$\upbeta$ line ratios are on average unchanged when fitting the data without rebinning.}.
To model the continuum, we use a library of simple stellar-population (SSP) spectra coupled with a 10\textsuperscript{th}-order multiplicative Legendre polynomial. The SSP spectra are the high-resolution (R=10,000) synthetic spectra from the C3K library \citep{conroy_stellar_halo_2019} with MIST isochrones \citep{choi_MIST_2016} and solar abundances. Note that repeating our fit with a polynomial background does not change our results.  
The emission lines and continuum are tied together to have the same velocity offset and velocity dispersion (while {\sc ppxf} also takes into account the variable spectral resolution). For
$[\mathrm{OII}]\lambda\lambda3727,3729$, we constrain the doublet ratio to its physical range; for
$\mathrm{[NeIII]}\lambda\lambda3870,3969$, we fixed the ratio to 0.31. As a sanity check, for
$[\mathrm{OIII}]\lambda\lambda4959,5007$, we leave the ratio a free parameter and check that the
recovered value is consistent within the errors with the theoretical expectation of 0.34 \citep[e.g.][]{osterbrock_astrophysics_2006}. We perform two fits: after the initial fit, we reject spectral pixels beyond four standard deviations from the best-fit model, then repeat the
procedure with outliers masked to obtain the final measurements.

The resulting fitted lines are shown in Fig.~\ref{fig:fit} (with zoomed insets for lines most relevant for this work) and the resulting line fluxes relative to H$\beta$ are given in Table~\ref{table:lines}. The uncertainties are taken from \textsc{ppxf} (up-scaled by $\sqrt{\chi^2}$), but we find them consistent with the values obtained from running one hundred Monte Carlo realisations of the best-fit spectrum, with random Gaussian noise taken from the noise vector.



\begin{figure*}
 \includegraphics[type=pdf,ext=.pdf,read=.pdf,width=1.\textwidth]{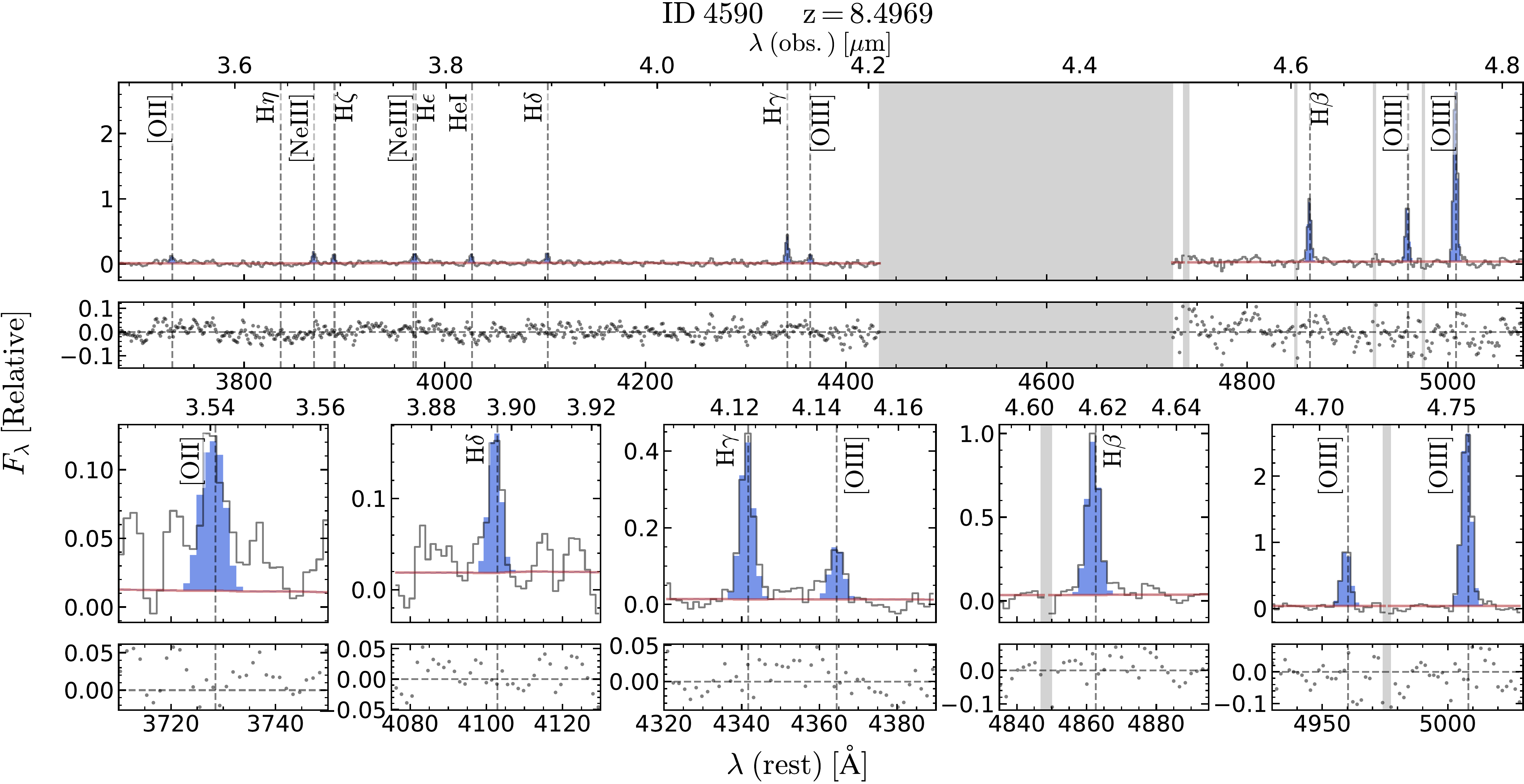}\\
 \vspace{0.35cm}
\includegraphics[type=pdf,ext=.pdf,read=.pdf,width=1.\textwidth]{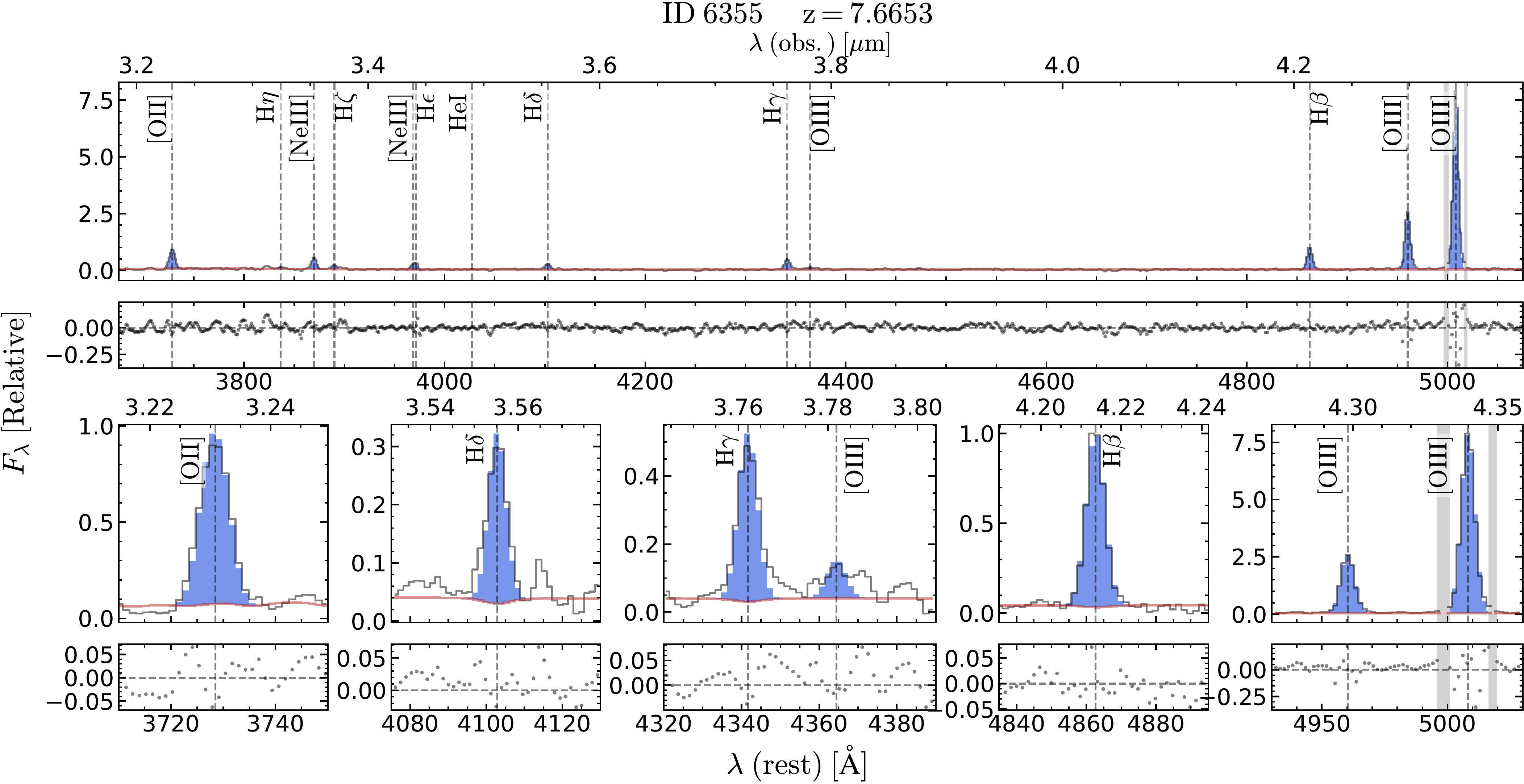}
  \caption{\textit{JWST}/NIRSpec spectra of the three sources analysed in this work. Several Hydrogen, Helium and metals emission lines are detected; the best-fit models are highlighted in blue. The data and best-fit continuum
  are traced by the solid grey and red lines, respectively; the dots are the residuals. The bottom panels
  show a zoom-in on the spectral region of three groups of lines; from left to right they are: the \oii\ doublet, H$\updelta$, H$\upgamma$ and [\ion{O}{iii}]4363, and H$\upbeta$ and \oiii. Grey regions have been masked due to artefacts (either in the spectrum or in the noise) or sigma clipping; the
  wide grey region at 4450-4750~\AA in the top panel of ID~4590 falls in the detector gap.}
  \label{fig:fit}
\end{figure*}

\setcounter{figure}{0}
\begin{figure*}
\vspace{0.1cm}
 \includegraphics[type=pdf,ext=.pdf,read=.pdf,width=1.\textwidth]{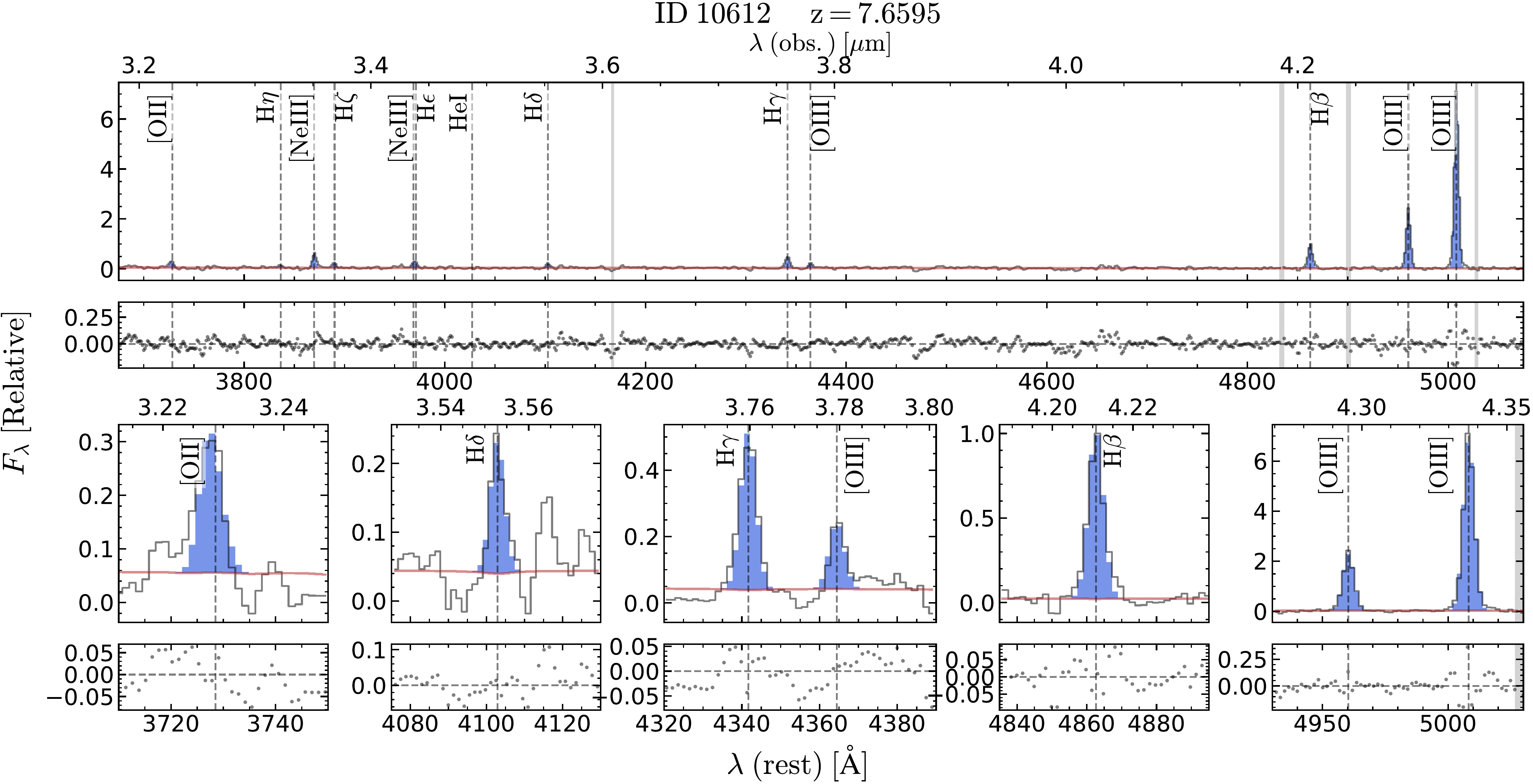}
  \caption{(continued)}\label{fig:fit1}
\end{figure*}


\begin{table*}
\caption{Observed line intensities normalised to H$\beta=1$ and errors on the ratio. The H$\beta$ emission line fluxes (in physical units of 10$^{-18}$~erg~s$^{-1}$~cm$^{-2}$) measured from the spectra are as follows: ID~4590~=$1.54\pm0.06$; ID~6355~=$2.11\pm0.05$; ID~10612~=$1.19\pm0.04$.}
\centering
\label{table:lines}
\begin{tabular}{@{}cccccccccc@{}}
\hline\hline
Galaxy ID & [\ion{O}{ii}]$\lambda\lambda3727,29$ & [\ion{Ne}{iii}]$\lambda3869$ & H$\zeta$ & H$\epsilon$ & H$\delta$ & H$\gamma$ & [\ion{O}{iii}]$\lambda4363$ & [\ion{O}{iii}]$\lambda4959$ & [\ion{O}{iii}]$\lambda5007$ \\
\hline
4590 & 0.15 $\pm$ 0.07 & 0.17 $\pm$ 0.03 & 0.13 $\pm$ 0.02 & 0.13 $\pm$ 0.02 & 0.16 $\pm$ 0.02 & 0.41 $\pm$ 0.03 & 0.14 $\pm$ 0.02 & 0.96 $\pm$ 0.05 & 3.08 $\pm$ 0.13 \\
6355 & 0.90 $\pm$ 0.10 & 0.45 $\pm$ 0.03 & 0.18 $\pm$ 0.02 & 0.18 $\pm$ 0.02 & 0.26 $\pm$ 0.02 & 0.46 $\pm$ 0.02 & 0.10 $\pm$ 0.02 & 2.66 $\pm$ 0.07 & 8.29 $\pm$ 0.21 \\
10612 & 0.26 $\pm$ 0.08 & 0.50 $\pm$ 0.04 & 0.19 $\pm$ 0.03 & 0.14 $\pm$ 0.03 & 0.17 $\pm$ 0.03 & 0.44 $\pm$ 0.03 & 0.18 $\pm$ 0.03 & 2.34 $\pm$ 0.09 & 7.11 $\pm$ 0.24 \\
\hline
\end{tabular}
\end{table*}

\subsection{NIRCam imaging}

\subsubsection{NIRCam observations}

We use the deep NIRCam imaging data of SMACS\,J0723.3–7327 from the Early Release Observations (Programme ID 2736) in the F090W, F150W, F200W, F277W, F356W, and F444W filters, which cover an observed wavelength range of $\lambda_{\rm obs}=0.8-5\micro\mathrm{m}$. We reduce the raw level-1 data products with the public \textit{JWST} pipeline (v1.6.1),\footnote{Available at \url{https://github.com/spacetelescope/jwst}.} using the latest available calibration files (CRDS\_CTX=jwst\_0927.pmap). An additional background subtraction is performed on the final mosaiced images by using \textsc{Photutils} to mask pixels identified with sources and then measuring the background with \textsc{photutils.Background2D}.

We conduct aperture photometry on the final mosaics with a range of aperture sizes, given the extended morphology of the three galaxies in this study and their close-by neighbours. Specifically, we use circular apertures with radii from $0.2-0.4$ arcsec. We perform an aperture correction with the help of WebbPSF \citep{2015ascl.soft04007P}.\footnote{Available at \url{https://github.com/spacetelescope/webbpsf}.} We find that the final SFRs and stellar masses only weakly depend on the aperture size. For simplicity, we therefore assume a fiducial size of 0.3 arcsec. We estimated the uncertainties of these fluxes from the error maps of the mosaic images.


\subsubsection{Spectral energy distribution fitting}

We perform SED modeling of the NIRCam photometry with the Bayesian code \textsc{beagle} \citep{2016MNRAS.462.1415C} with the main aim to derive stellar masses and SFRs for our three galaxies. For consistency with the local MZR and FMR considered here (discussed in Section~\ref{section:Evolution_of_the_metallicity_scaling_relations}), we use a \citet{2003PASP..115..763C} stellar initial mass function 
(IMF).\footnote{Arguably, this IMF may not be appropriate for these young, low-metallicity galaxies. However, it is not established what the correct IMF at such early times should be and an IMF conversion only introduces a constant scaling factor.} For the star-formation history (SFH), we assume a delayed-exponential form. We find for all three galaxies stellar masses and SFRs of $10^{8-9}~\mathrm{M_{\odot}}$ and $16-65~\mathrm{M_{\odot}}~\mathrm{yr}^{-1}$ (Table~\ref{table:properties}). Furthermore, the inferred stellar (and gas-phase\footnote{In SED modeling, we assume that the stellar and gas-phase metallicity are the same.}) metallicities are significantly sub-solar (12+log(O/H)=$7.25\pm0.21$, $7.53\pm0.08$, and $7.50\pm0.12$) and an overall low attenuation ($A_V$=$0.37\pm0.04$, $0.50\pm0.03$ and $0.21\pm0.03$) for ID$4590$, ID$6355$ and ID$10612$, respectively. It is interesting that we obtain broad-band metallicities fairly consistent with the \Te ones but for ID$6355$ (which is underestimated by a factor of $\sim5$), confirming in particular the very low metallicity of ID$4590$ (see section~\ref{sec:oxygen_abund}).

Within the \textsc{beagle} framework, we also explore a constant SFH, which has however little effect on the inferred stellar masses and SFRs. Finally, we explore two other SED fitting codes, namely \textsc{bagpipes} \citep{2018MNRAS.480.4379C} and \textsc{prospector} \citep{2021ApJS..254...22J}. We run \textsc{prospector} with the same setup as in \citet{2022ApJ...927..170T}. With \textsc{bagpipes}, we explored three alternative SFHs: either constant, delayed exponential or delayed exponential with burst. Reassuringly, we find that the derived stellar masses and SFRs are consistent within $1 \sigma$ uncertainties between the different codes, as well as with those obtained by \cite{tacchella_stellar_pop_2022} via a \textsc{prospector} run including emission lines and assuming a bursty prior on the SFH.
Moreover, the stellar masses (converted to the same IMF) agree within $1 \sigma$ uncertainties with the results presented by \citet{2022arXiv220708778C}.
However, a more thorough comparison between these different SED fitting codes goes beyond the scope of the present paper.

\subsubsection{Magnification factors}


The galaxies analysed in the present paper are background galaxies of the SMACS\,J0723.3–7327 lensing cluster with publicly available lens models. 
In this work, we exploit the models recently provided by \cite{mahler_lensmodel_smacs_2022}, which combine ancillary \textit{HST} with novel \textit{JWST}/NIRCAM data to better constrain the cluster mass distribution.
Magnification maps are derived for each target redshift, and the average value from within a 1\arcsec-wide box around the central coordinates of each galaxy is assumed as the fiducial magnification factor. The associated uncertainty is computed from the standard deviation of one hundred realisations of the magnification maps obtained through Monte Carlo simulations by perturbing the input model parameters.
These results in the magnification factors listed in Table~\ref{table:properties}.

Additional models for this cluster have been recently published by
\cite{Pascale2022} and \cite{Caminha2022}. A full assessment of the systematics introduced by the adoption of different lensing models is far beyond the scope of this work. However we remark that, for the highest magnification galaxy ID$4590$, the different values obtained by the available models can impact the stellar mass and SFR determination for this source up to a factor of $30-50\%$.

\section{Line ratios and abundances at \texorpdfstring{$z \sim$ 8}{z~8}}
\label{sec:line_ratios}

\subsection{Balmer decrements and dust attenuation}



\begin{figure}
    \centering
  \includegraphics[width=\columnwidth]{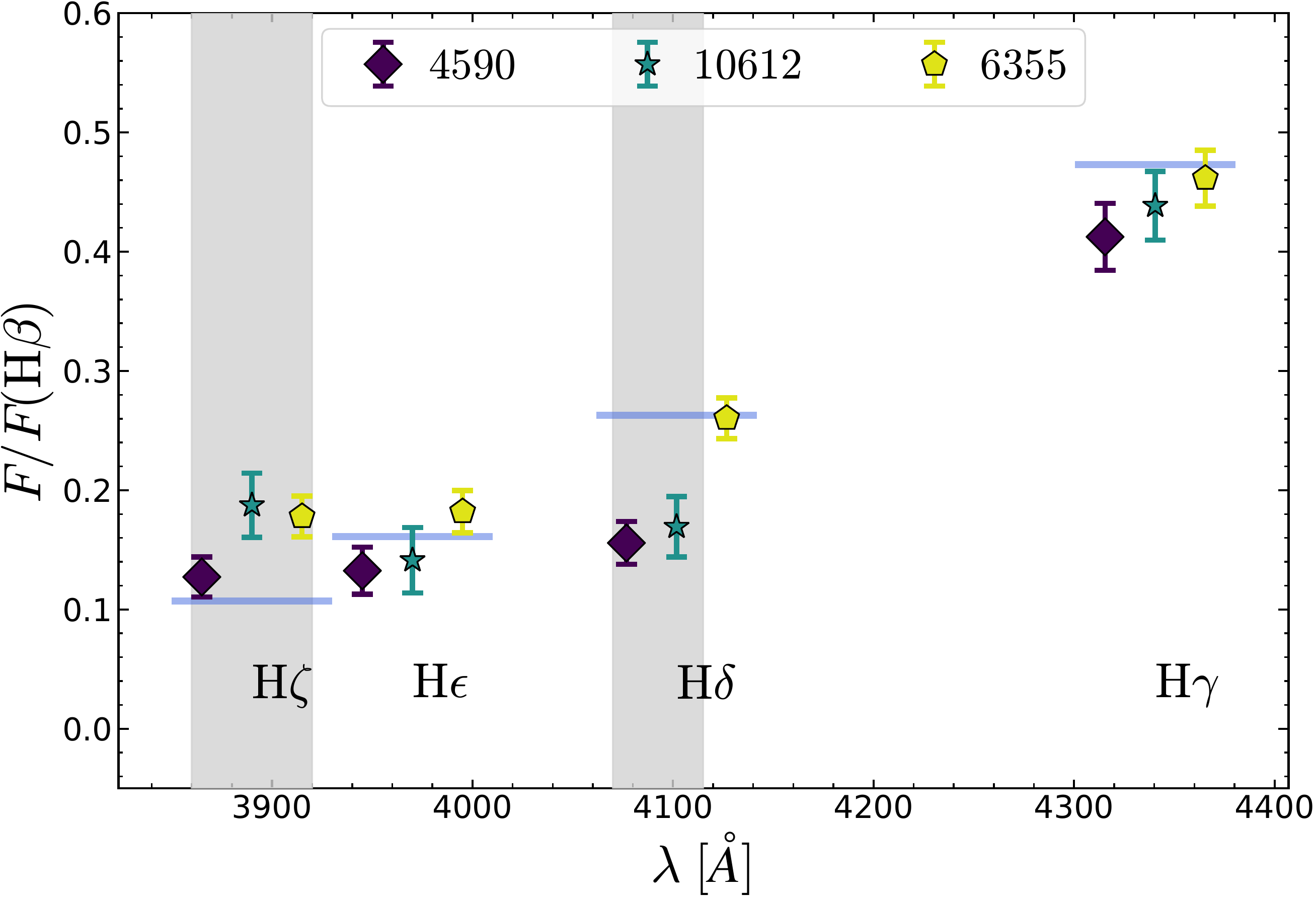}
    \caption{Flux ratios between different Balmer lines and the H$\beta$ as observed in all galaxies' spectra. The theoretical values expected from atomic physics (assuming the case B recombination at T$=1.5\times10^{4}$K and N$_{\rm e}$ = 300 cm${-3}$) are marked by the blue lines. All observed ratios are consistent with little dust attenuation. Grey shaded areas mark regions with unreliable ratios: non-physical H$\zeta$/H$\beta$ ratio might be driven by H$\zeta$ flux being contaminated by He I emission, whereas residuals in the background subtraction may affect the H$\delta$ emission in ID$4590$ and ID$10612$.
    }
    \label{fig:balmer_decrements}
\end{figure}

Fig.~\ref{fig:balmer_decrements} shows a comparison between the observed and theoretical fluxes relative to H$\beta$ for different Balmer emission lines, namely H$\zeta$, H$\epsilon$, H$\delta$, and H$\gamma$.
The theoretical values of the corresponding Balmer decrements computed using PyNeb \citep{luridiana_pyneb_2012,luridiana_pyneb_2015}, assuming Case B recombination, an 
electronic temperature $\Te=1.5\times10^{4}$~K and an electron density $\Ne=300$\,cm$^{-3}$ 
(i.e., H$\upzeta$/H$\upbeta\approx0.105$, H$\upepsilon$/H$\upbeta\approx0.159$, 
H$\updelta$/H$\upbeta\approx0.259$, H$\upgamma$/H$\upbeta\approx0.468$), are marked by the 
horizontal blue lines. Under realistic assumptions, these ratios are fairly insensitive to the precise physical conditions of the ISM. For example, varying the temperature between \Te$= 0.5\times 10^4$~K and \Te$= 3\times 10^4$~K and the density 
between $\Ne = 10$~cm$^{-3}$ and $\Ne = 10,000$~cm$^{-3}$ would make these ratios change by at 
most $\sim 3.5$~per\ cent.


We note that while a poorly corrected stellar absorption could potentially bias the Balmer ratios, the continuum is nearly undetected in the spectra of these three galaxies. Moreover these are extremely young galaxies that are not expected to have significant Balmer absorption based on the level of continuum detection. While Balmer-line absorption is expected to be weaker for H$\upbeta$ than for H$\upgamma$ and H$\updelta$ at least, even assuming an equivalent width of 4~\AA\ for H$\upbeta$ absorption and 0~\AA\ for the other Balmer lines would lead to a correction of less than 5~per\ cent. 


The Balmer ratios shown in Fig.~\ref{fig:balmer_decrements} are consistent with the theoretical value, or slightly lower. The exceptions are H$\upzeta$ for galaxies ID$10612$ and ID$6355$, however, H$\zeta$ is very faint and possibly contaminated by HeI. The other exception is H$\delta$, which is significantly lower than the theoretical value for galaxies ID$4590$ and ID$10612$, while is well in agreement with the theoretical value for ID$6355$; however, the former two galaxies are those with the lowest S/N and, in addition, some background residual problems seem to be present at these wavelengths for these galaxies. We note that, as already reported by \cite{Schaerer22}, the use of the level 3 products from the MAST archive 
would result in Balmer ratios much higher than the theoretical values, confirming the presence of residual issues in the data processing and/or flux calibration suggested both in the JWST documentation\footnote{see \hyperlink{https://jwst-docs.stsci.edu/jwst-calibration-pipeline-caveats}{https://jwst-docs.stsci.edu/jwst-calibration-pipeline-caveats}} and by the significant levels of spectrophotometric correction we had to introduce (see Sect.~\ref{sec:data_reduction}).

We simultaneously fit the ratios of H$\upgamma$/H$\beta$ and H$\upepsilon$/H$\beta$ to infer the dust attenuation by assuming the SMC curve from \cite{gordon_LMC_attenuation_2003} with $R_V=2.505$; as mentioned, H$\updelta$/H$\beta$ appears strongly underestimated in both ID$4590$ and ID$10612$ compared to the other two Balmer decrements at fixed attenuation, possibly due to poor continuum subtraction, hence we consider this ratio only for ID$6355$.
The inferred values of nebular $A_V$ for the three sources are reported in Table~\ref{table:properties}.
The use of the locally-derived \cite{gordon_LMC_attenuation_2003} SMC curve may be arguable for such distant galaxies, however it as been shown to be adequate for $z\sim2$ low-metallicity galaxies \citep{shivaei_mosdef_2020}.
Moreover, since the [\ion{O}{iii}]$\lambda4363$ emission line is close in wavelength to the H$\gamma$ line and \oiii\ is close to H$\upbeta$, and considering that the attenuation is mostly driven by H$\upgamma$/H$\upbeta$, the choice of the attenuation curve has a minor effect on the dust-corrected value of the [\ion{O}{iii}]4363/[\ion{O}{iii}]5007 line ratio (and hence on the inferred temperature and metallicity).

We compare the nebular attenuation to the attenuation of the continuum derived  from the \textsc{beagle} SED fitting, once scaling stellar to nebular attenuation by a factor of $2.11$\footnote{we note the exact determination of this factor is quite uncertain}, following \cite{shivaei_mosdef_2020}.
We find agreement within the uncertainties for ID$4590$ and ID$10612$, whereas a significantly higher attenuation from the SED fitting in ID$6355$. 

\subsection{Excitation diagnostics}



\begin{figure*}
    \centering
  \includegraphics[width=0.48\textwidth]{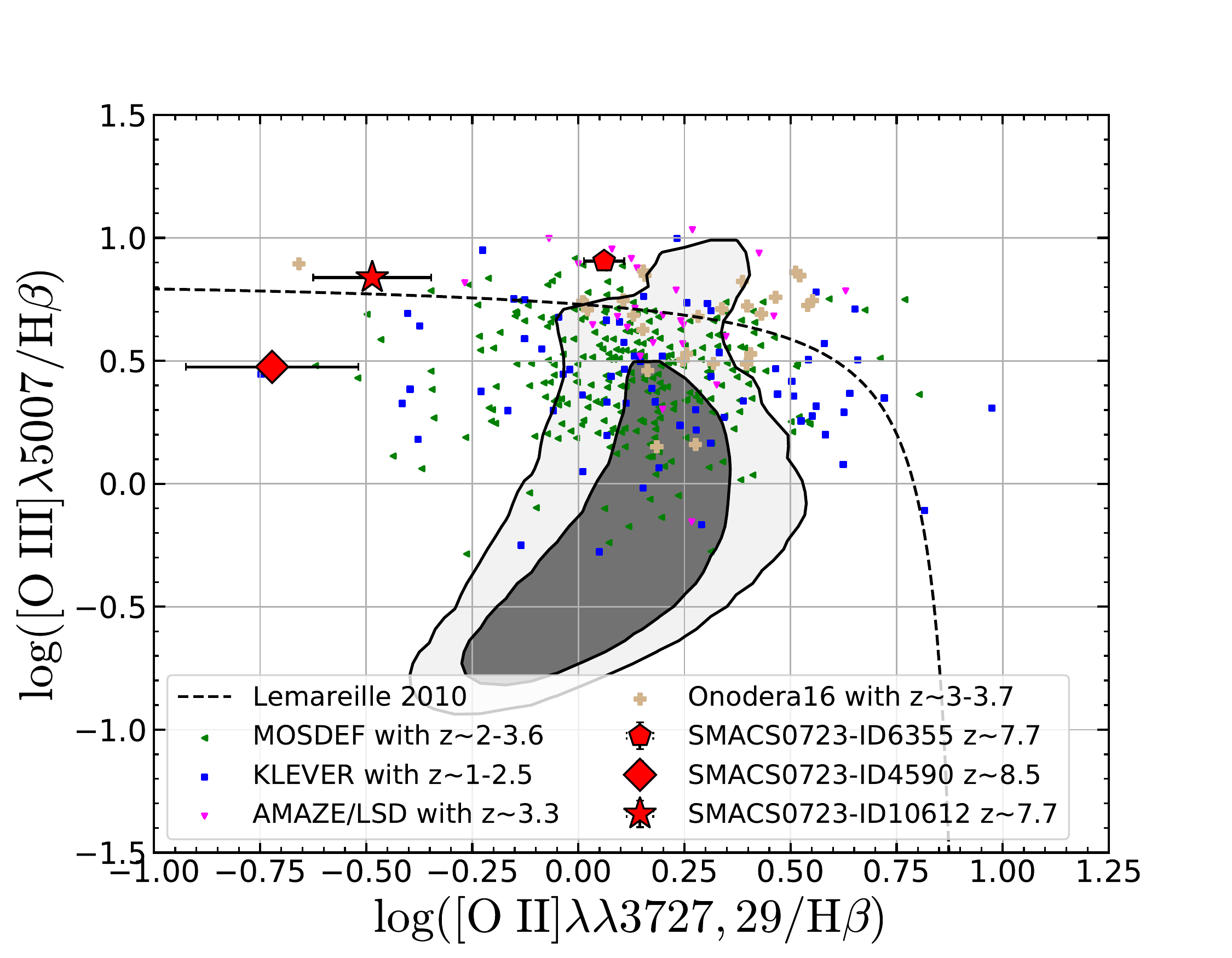} 
  \includegraphics[width=0.48\textwidth]{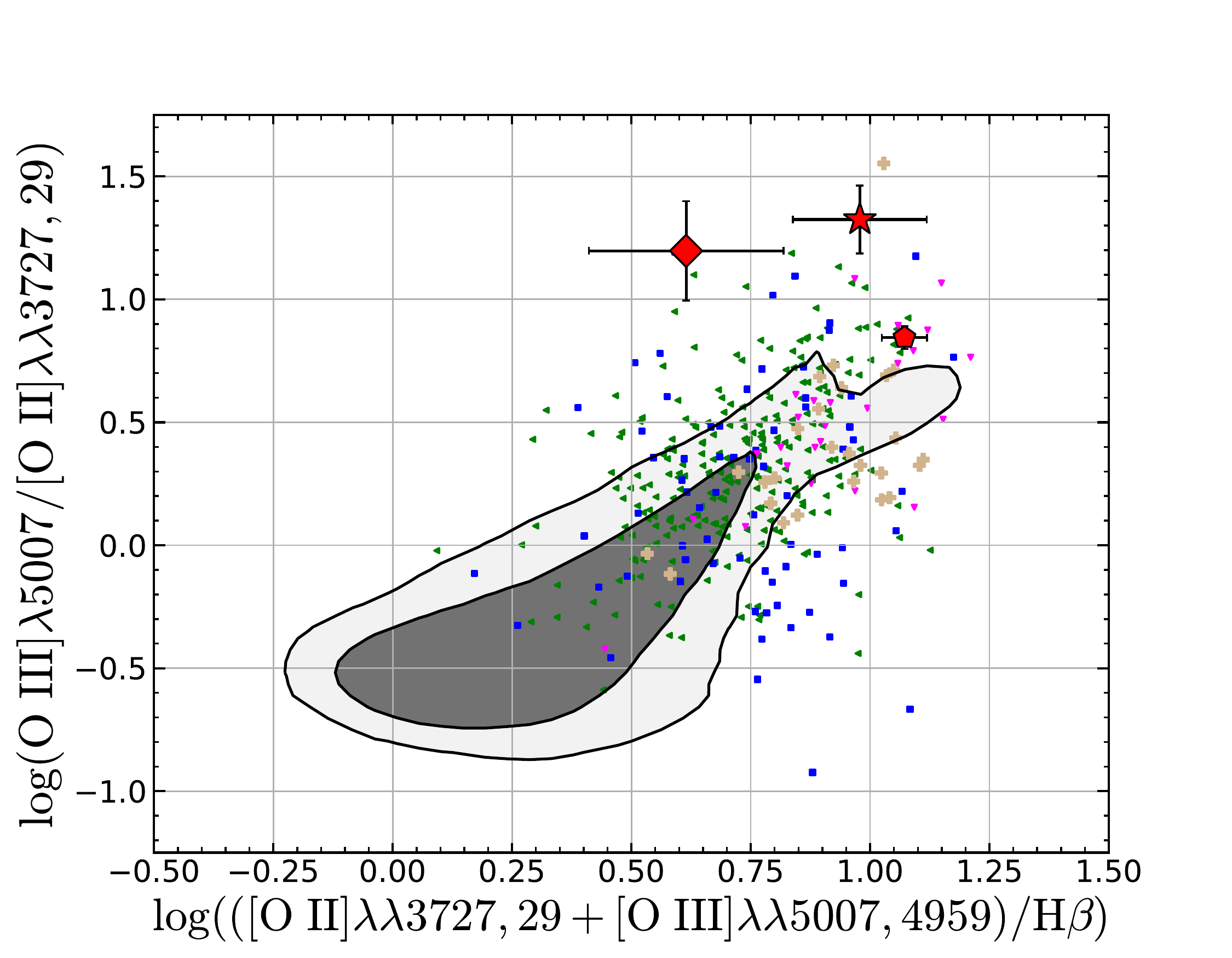}

    \caption{[\ion{O}{iii}]/H$\beta$ versus [\ion{O}{ii}]/H$\beta$ diagram (left-hand panel) and  [\ion{O}{iii}]/[\ion{O}{ii}] versus R$_{23}$ diagram (right-hand panel) for local SDSS galaxies (grey contours marking the 70\% and 90\% of the distribution), galaxies at $z\sim$1--3 from the literature, and the \textit{JWST} sample at $z\sim8$.}
    \label{fig:R3R2}
\end{figure*}

At the redshifts of our targets H$\alpha$ and adjacent nebular lines are shifted outside the NIRSpec band, so they cannot be used to construct the classical BPT diagnostic diagrams \citep{baldwin_classification_1981}. However, interesting information about the excitation properties of these sources can be inferred from the \oiii/H$\beta$ versus \oii/H$\beta$ diagram. This diagram is shown in the left-hand panel of Fig.\ref{fig:R3R2}, where the shaded region marks the distribution of local galaxies in the MPA-JHU catalog from the Sloan Digital Sky Survey \citep{tremonti_origin_2004, brinchmann_physical_2004} and other surveys at $z\sim1-3$ \citep{Troncoso2014, onodera_ism_2016, sanders_mosdef_mzr_2021, hayden-pawson_NO_2022}. The dashed line indicates the local dividing line between star-forming galaxies and active galactic nuclei (AGN) identified by \cite{Lamareille2010}.

Intermediate-redshift star-forming galaxies are characterized by a large scatter, but they also tend to be shifted towards higher values of [\ion{O}{iii}]/H$\beta$. This is an effect similar to that seen in the classical BPT diagram \citep[see the discussion in][]{maiolino_re_2019}, which has been attributed to a combination of higher ionization parameter and harder radiation field associated with $\alpha$-enhanced stellar populations \citep[e.g.][, but see also \citealt{Curti22}]{Strom18,topping_mosdef-lris_2020_i}. Therefore, finding that the two galaxies at $z\sim$~7.6 are located above \cite{Lamareille2010}'s line does not necessarily imply that these are AGN, but that they follow the same trend as intermediate-redshift galaxies. 

\begin{figure}

    \centering
  \includegraphics[width=0.98\columnwidth]{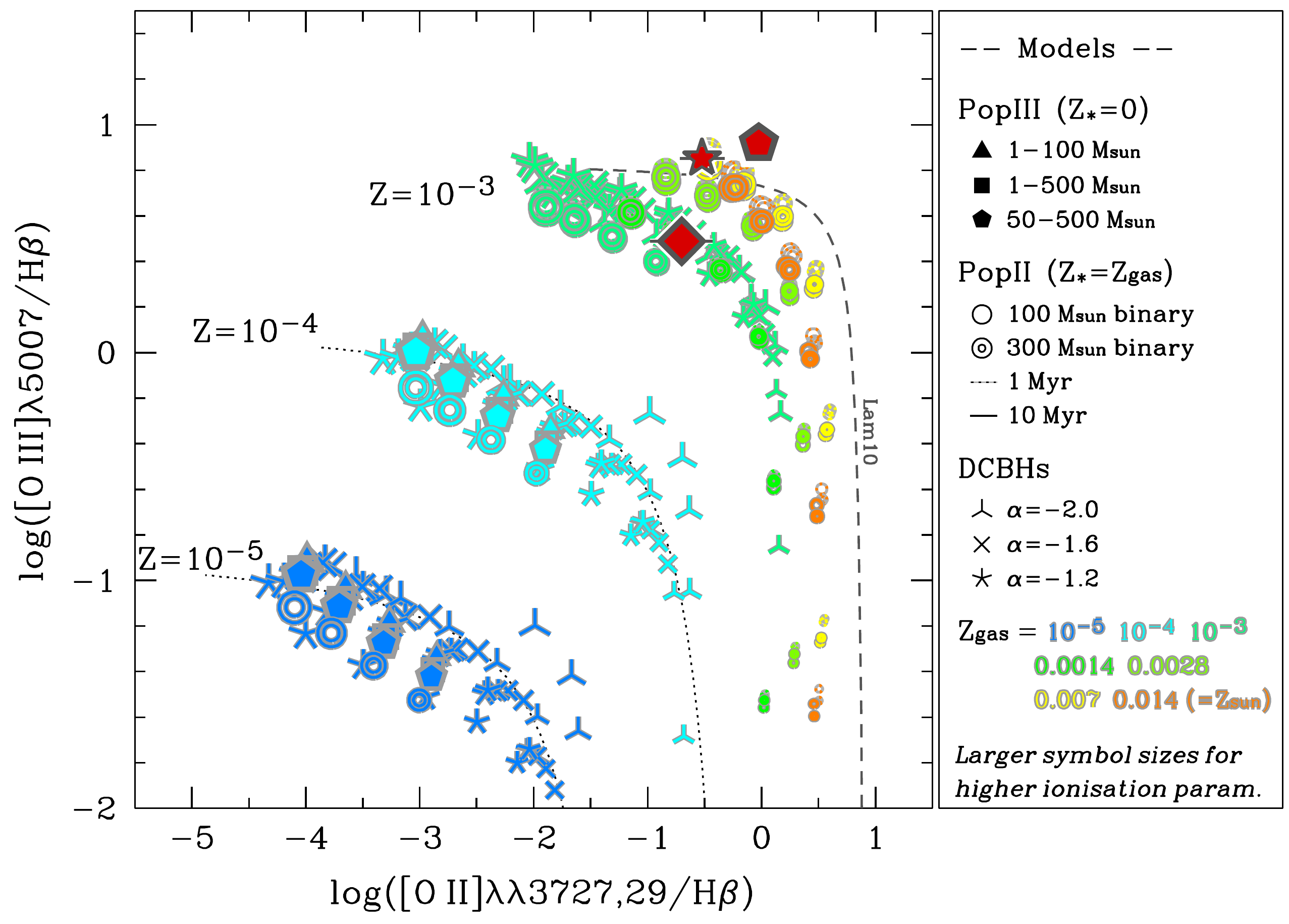} 

    \caption{[\ion{O}{iii}]/H$\beta$ vs [\ion{O}{ii}]/H$\beta$ diagram illustrating the photoionization models
	for low metallicity PopII, PopIII and AGNs from \citealt{Nakajima_Maiolino22}. Different symbols mark different population of objects (as explained in the legend), while different colours belong to different values of the gas-phase metallicity. 
	The line ratios observed in our $z\sim$~8 galaxies are marked with red symbols. 
	}
    \label{fig:nakajima_diagram}
\end{figure}

Galaxy ID$4590$ at $z\sim8.5$ is located in a region of the diagram poorly populated by both local and intermediate-redshift galaxies, with log([\ion{O}{iii}]/H$\beta$)$\sim$~0.5 and log([\ion{O}{ii}]/H$\beta$)$\sim$~-0.75. 
According to the photoionisation models presented in
\cite{Nakajima_Maiolino22}, and shown in Fig.~\ref{fig:nakajima_diagram}, this region is populated by young, metal poor galaxies. 
However, we also note that the presence of an AGN with low metallicity  ($Z\sim10^{-3}$) can not be completely ruled out by these models.

The right-hand panel of Fig. \ref{fig:R3R2} shows instead the excitation diagram O$_{32}$=\oiii/\oii\ versus R$_{23}=$([\ion{O}{iii}]$\lambda4959,5007$ + [\ion{O}{II}]$\lambda3727,29$)/H$\beta$, where the same local and intermediate-redshift galaxies are shown as in the left-hand panel . This diagram can be considered as a proxy of ionization parameter (primarily traced by O$_{32}$, \citet{diaz_hii_regions_2000}) versus metallicity (primary traced by R$_{23}$, e.g. \citet{nagao_gas_2006}). However, high O$_{32}$ values are sometimes also considered as indicative of density-bounded clouds, possibly associated with high escape fractions ($f_{\text{esc}}$) of Lyman-continuum (LyC) photons \citep{Nakajima14,Nakajima20,Barrow2020}.
All three galaxies at z$>$7 considered here have very high values of O$_{32}$. Especially, ID$10612$ at z=7.6 and ID$4590$ at z=8.5 have extremely high values of O$_{32}$ that are rarely seen even in galaxies at z$\sim$2--3, if not in extreme \oiii\ emitters with very high equivalent widths \citep{tang_2019}.
The fact that these galaxies have such high values of O$_{32}$ can potentially indicate very high ionisation parameters and/or large LyC escape fractions \citep[]{Chisholm_escape_frac_2022}.
Based on Ly$\upalpha$ emitters (LAEs) probed by different low- and intermediate-redshift galaxy surveys with similar line ratios and metallicities to our JWST sample we can provide (following e.g., \citealt{Nakajima20,Izotov2018a}) a rough estimate of the $f_{\text{esc}}$ in our galaxies. In particular, the O$_{32}$ value for ID6355 ($\log\,\mathrm{O}_{32}=0.96$) is consistent with LyC leakers with $f_{\text{esc}} \lesssim 0.1$, whereas ID4590 and ID10612 show O$_{32}$ values ($\log\,\mathrm{O}_{32}=1.18$; $1.36$, respectively) comparable with $f_{\text{esc}}$ from $\sim0.1$ up to $\sim0.5$. We stress however that the relationship between $\mathrm{O}_{32}$ and $f_{\text{esc}}$ is quite scattered, and that a large $\mathrm{O}_{32}$ may be a necessary but not sufficient condition for the presence of relevant LyC emission \citep{izotov_2017}.

\subsection{Oxygen abundance determination}
\label{sec:oxygen_abund}
We exploit the highly significant detection of the [\ion{O}{iii}]$\lambda4363$ auroral line in \textit{JWST}/NIRSpec spectra of the three galaxies studied here to measure the oxygen abundance (a proxy of the total gas-phase metallicity) by means of the electron-temperature (\Te) method. More specifically, [\ion{O}{iii}]$\lambda4363$ is detected at $8.6\sigma$, $7.3\sigma$, and $6.4\sigma$ in the spectra of ID$4590$, ID$10612$, and ID$6355$, respectively.

For the purposes of temperature and abundance computation, we model each galaxy as a single \Hii region constituted by two ionisation zones: the high-ionization one, traced by the O$^{++}$ ion; and the low-ionization one, traced by the O$^{+}$ ion.
We have direct access to the temperature of the high-ionisation zone (t3) from the ratio between the [\ion{O}{iii}]$\lambda4363$ and [\ion{O}{iii}]$\lambda5007$ emission lines.

We implement the iterative procedure described in \cite{nicholls_measuring_2013} to infer t3 from this auroral-to-nebular diagnostic line ratio, adopting the atomic data from \cite{palay_improved_2012}, and assuming an electron density $\Ne=300$\,cm$^{-3}$ (a value typically measured in the ISM of $z\sim 2-3$ galaxies, e.g. \citealt{sanders_mosdef_2016}); different values of \Ne\ would not impact significantly our results, as the [\ion{O}{iii}]$\lambda4363$/[\ion{O}{iii}]$\lambda5007$ ratio depends only mildly on gas density. 
The derived t3 temperatures for our galaxies are reported in Table~\ref{table:properties}.
The temperatures derived for ID$10612$ and ID$6355$ ($\sim17600$~K and $\sim12,200$~K) are consistent with the typical values of low- and intermediate-metallicity \Hii regions, whereas the ID$4590$ source at $z\sim8.5$ is characterised by a much higher temperature ($\sim28,600$~K), possibly driven by low metallicity and high ionisation parameter, and similar to what observed in local extremely metal-poor galaxies \citep[e.g.][]{Izotov2018,Izotov2021}.

Since we do not have access to temperature diagnostics for the low-ionisation species (i.e., no auroral line emission is detected from N$^{+}$ or S$^{+}$ species in the spectra), here we adopt the temperature-temperature relation from \cite{pilyugin_electron_2009} to infer the temperature of the low-ionisation zone (t2): $t2 = 2640 + 0.835\times t3$.
While the applicability of such relation at $z\sim 8$ is questionable, adopting a different locally calibrated t2-t3 relation does not significantly impact the total oxygen abundance determination since in these low-metallicity environments the O$^{++}$ abundance is largely dominant over O$^{+}$ \citep{andrews_mass-metallicity_2013, curti_new_2017}.
For comparison, adopting the t2-t3 relation from \cite{garnett_nitrogen_1990} impacts the final metallicity calculation by less than 1~per\ cent.


Finally, we compute the ionic abundances of O$^{+}$ and O$^{++}$ using the \textit{getIonAbundance} routine of \textsc{pyneb} \citep[v 1.1.10,][adopting the atomic data from \citealt{palay_improved_2012}]{luridiana_pyneb_2012,luridiana_pyneb_2015}, which provides the abundance of ionic species given the electron temperature (of the corresponding ionisation zone), the gas density, and the flux of a strong emission line (e.g., \oiii\ for O$^{++}$, \oii\ for O$^{+}$) relative to H$\beta$., and assuming the same atomic data from \cite{palay_improved_2012}.
Here, we assume that the total oxygen abundance is the sum of the abundances of the singly- and doubly-ionised oxygen species, i.e., $\text{O/H} = \text{O}^{+}/\text{H}\ +\ \text{O}^{++}/\text{H}$, neglecting contributions from higher ionisation states. 


The derived electron temperatures and oxygen abundances for our galaxy sample are reported in Table~\ref{table:properties}; uncertainties on both quantities are computed through Monte Carlo simulations by randomly perturbing (assuming a Gaussian noise distribution) one hundred times all measured line fluxes by their measurement errors\footnote{we verified that errors from the \textsc{ppxf} fitting procedure are consistent with those inferred from bootstrapping} (including the uncertainty from reddening correction), randomly varying the density between $100$ and $1000$~cm$^{-3}$, and repeating the full temperature and abundance calculation for each iteration, taking the standard deviation of the distribution of temperatures and abundances obtained as the representative uncertainties on the measured values. 
Nonetheless, we note that these values probably underestimate the true uncertainties, because the systemic contributions are not considered. In particular, temperature stratification in \Hii regions can significantly halter the derived metallicity \citep{stasinska_abundance_2002,kewley_understanding_2019}.

As briefly mentioned above, the high t3 measured in ID$4590$ places this object at the limit of what has ever been observed in the local Universe for star-forming galaxies.
Such high temperature might also reveal the presence of a much harder ionising continuum associated with AGN activity \citep{Dors2020, Riffel2021}, which is indeed not ruled out also by photoionisation models, as already shown by comparing the observed line ratios with the predictions of the \cite{Nakajima_Maiolino22} models in Fig.~\ref{fig:nakajima_diagram}.   
Although we favour the scenario of star-forming galaxy with low metallicity and high ionisation parameter, for comparison we verified that the analytic approach presented in \citet[][equations 9 and 10]{Dors2020}, tuned for deriving \Te-based metallicities in AGNs, provides a comparable estimate of the total oxygen abundance, finding temperatures and metallicities consistent within the uncertainties with our fiducial \Te-method.
We simply note that, in case of the presence of an AGN, the metallicity of ID$4590$ might be underestimated by not accounting for the abundance of the O$^{+++}$ species, and that the stellar mass and SFR determination would require further care in decomposing the stellar from the AGN contribution to the photometry.
However, we also note that obscuration from the dusty tori in type II AGNs could weaken the [\ion{O}{iii}]$\lambda4363$ emission \citep[e.g][]{Nagao2001}.

It is also interesting to note that the galaxy with the lowest metallicity, ID$4590$, is also the one with the highest dust attenuation. While one should not over-interpret this result, as the uncertainty on attenuation is large, this is something that is observed in some local dwarf, low metallicity galaxies (e.g. SBS 0335-052), which can have large dust masses and high dust attenuation. At a given gas mass, compactness and geometrical effect can strongly contribute to the higher dust attenuation; moreover, high gas density can foster the rapid growth of dust in the ISM \citep{Schneider16}.

Finally, we note that, in a recent paper, \cite{Schaerer22} analyses the same three galaxies, obtaining different results in terms of gas temperature and metallicity.
They notice non-physical Balmer decrements in their spectra, and apply a power-law, wavelength dependent correction. 
They also infer un-physically high temperature for ID$4590$.
As discussed in the previous sections, we obtain reasonable Balmer decrements, without introducing any {\it a posteriori} adjustment, and we can reliably measure the temperature in ID$4590$, which results consistent with the ISM properties of extremely metal-poor galaxies. 
We believe that these differences can be ascribed to our additional processing of the data, as discussed in detail in Sect.~\ref{sec:data_reduction}. 



\begin{table}
\caption{Derived galaxy properties.}
\centering
\label{table:properties}
\begin{tabular}{@{}lccc@{}}
\hline\hline
Galaxy ID                               &        4590    &         6355     &       10612 \\
\hline \vspace{0.05cm}
Redshift    &                             8.496    &        7.665      &      7.658 \\
$\mu^{a}$                         & 3.74 $\pm$ 0.07 &   1.231 $\pm$ 0.002 &  1.339 $\pm$ 0.003 \\
log(M$_{\star}$/M$_{\odot}$)$^{b}$  & 7.75 $\pm$ 0.07 &   8.72 $\pm$ 0.04 &  8.08 $\pm$ 0.04 \\
log(SFR [M$_{\odot}$yr$^{-1}$])$^{b}$    & 0.35 $\pm$ 0.07 &   1.47 $\pm$ 0.04 &  0.90 $\pm$ 0.04 \\
$A_V$ [nebular]$^{c}$                 & 0.68 $\substack{+0.34 \\ -0.25}$  &  0.0 $\substack{+0.1\\0.0}$ &  0.40$\substack{+0.46 \\ -0.27}$  \\
$A_V$ [stellar]$^{d}$                      & 0.37 $\pm$ 0.04 &   0.50 $\pm$ 0.03 &  0.21 $\pm$ 0.03 \\
\Te([\ion{O}{iii}])[$10^{4}$~K]             & 2.77 $\pm$ 0.42 &   1.20 $\pm$ 0.07 &  1.75 $\pm$ 0.16 \\
12+log(O/H)                   & 6.99 $\pm$ 0.11 &   8.24 $\pm$ 0.07 &  7.73 $\pm$ 0.12 \\

\hline
\end{tabular}
{\raggedright 

$^{a}$ derived from the lens models presented in \citealt{mahler_lensmodel_smacs_2022} \newline
$^{b}$ values are corrected for magnification (errors on $\mu$ are propagated) \newline
$^{c}$ inferred from ratios of Balmer lines \newline
$^{d}$ inferred from SED fitting
\par}
\end{table}

\begin{figure*}
    \centering
  \includegraphics[width=0.9\textwidth]{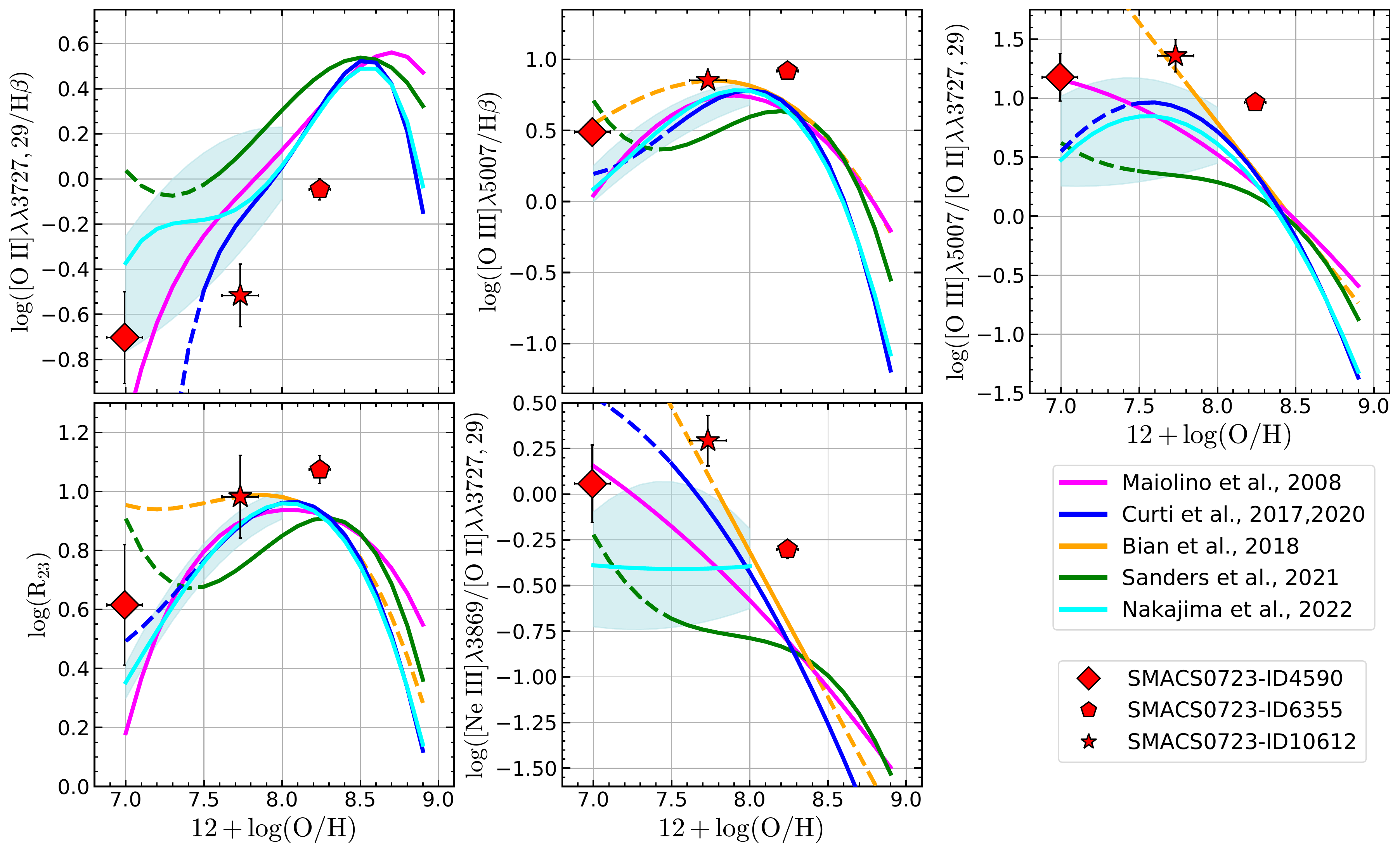}
    \caption{The relationships between \Te metallicity and strong-line ratios for the \textit{JWST} galaxy sample are compared with some widely adopted abundance calibrations, namely from \citealt{maiolino_amaze_2008}, \citealt{bian_ldquodirectrdquo_2018}, \citealt{curti_new_2017, curti_massmetallicity_2020}, \citealt{sanders_mosdef_mzr_2021}, and \citealt{nakajima_empress_2022}. All these calibration relations are built from samples of local star-forming galaxies with \Te-based measurements, but are based on slightly different assumptions (see the text for more details). Solid lines represent the relationships in the metallicity calibration range as provided by the authors, whereas dashed lines mark extrapolations of the polynomial fit outside that range. Shaded areas encompass the region between the calibrations based on the low- and high-EW(H$\beta$) sub-samples of metal-poor galaxies as presented in  \citealt{nakajima_empress_2022}. }
    \label{fig:calibrations}
\end{figure*}

\subsection{Comparison with strong-line metallicity calibrations}

Metallicity diagnostics based on strong nebular emission-line ratios are widely adopted by the astronomical community to infer the chemical abundances of large samples of galaxies, for which auroral lines are often not detected.
Several authors have calibrated such diagnostics by exploiting samples of local star-forming galaxies and \Hii regions with \Te-based metallicity determinations \citep[e.g.][]{pettini_oiiinii_2004,maiolino_amaze_2008,marino_o3n2_2013, pilyugin_new_2016, curti_new_2017, bian_ldquodirectrdquo_2018, sanders_mosdef_mzr_2021}.
However, an evolution in the ISM conditions of high-redshift galaxies compared to the local Universe might impact the intrinsic dependence of strong-line ratios on gas-phase metallicity, potentially hampering their use as abundance diagnostics at high redshift and thus biasing the assessment and interpretation of the chemical evolution history of galaxies.

Here, we exploit the \Te-based abundances delivered by \textit{JWST}/NIRSpec spectra to test the rest-frame optical, strong-line metallicity calibrations for the first time at $z\sim 8$. 
In Fig.~\ref{fig:calibrations}, the \Te-based log(O/H) measurements of the three galaxies in our sample are plotted against some of the most widely adopted strong-line metallicity diagnostics, specifically: R3=\oiii/H$\upbeta$, R2=\oii/H$\upbeta$, R$_{23}$=([\ion{O}{iii}]$\lambda\lambda4959,5007$+\oii)/H$\upbeta$, O32=\oiii/\oii, and Ne3O2=[\ion{Ne}{iii}]$\lambda3869$/\oii.

Different calibration curves (in the form of polynomial relations between metallicity and strong-line ratios) are shown for these diagnostics in each panel, namely from \cite{maiolino_amaze_2008} \cite{curti_massmetallicity_2020}, \cite{sanders_mosdef_mzr_2021}, \cite{bian_ldquodirectrdquo_2018} and \cite{nakajima_empress_2022}. In brief, \cite{curti_massmetallicity_2020} complemented the \cite{curti_new_2017} calibrations based on \Te metallicity measurements performed on stacked spectra of SDSS galaxies in the \oii/H$\upbeta$ versus \oiii/H$\upbeta$ diagram; here, we also include a new calibration of the Ne3O2 diagnostic based on the same methodology and combined sample of stacks and individual SDSS galaxies with [\ion{O}{iii}]$\lambda4363$ detections described in \cite{curti_new_2017}, and which was not published before.\footnote{The calibration is presented in the usual form [log(R)=$\sum_{i=0}^n a_ix^i$], where x=12+log(O/H)-8.69 (solar metallicity from \citealt{allende_prieto_forbidden_2001}), and the best-fit coefficients are: $c_0=-1.632$;\ $c_1=-2.061$;\ $c_2=-0.461$.}  
\cite{sanders_mosdef_mzr_2021} combined the measurements performed by \cite{andrews_mass-metallicity_2013} in stellar mass (\mstar)-SFR stacks of SDSS galaxies with a sample of low-metallicity dwarfs from \cite{berg_direct_2012}, whereas \cite{maiolino_amaze_2008} combined a sample of galaxies with auroral line detections at low metallicity from \cite{nagao_gas_2006} with predictions from photoionisation models in the high-metallicity regime. \cite{nakajima_empress_2022} instead extended the calibrations based on the \cite{curti_new_2017} SDSS stacks to the low-metallicity regime leveraging \Te-measurements in extremely metal-poor galaxies from the EMPRESS survey \citep{kojima_2020_empress}; in particular, in Fig.~\ref{fig:calibrations} we show the calibrations best-fit obtained by \cite{nakajima_empress_2022} from the full galaxy sample with the solid line, whereas the shaded areas encompass the region between the calibrations based on the sub-samples of metal-poor galaxies characterised by high- and low-EW(H$\beta$) (i.e., EW(H$\beta$)~$>200$~\AA\ and $<100$~\AA, respectively).
Finally, \cite{bian_ldquodirectrdquo_2018} built a set of calibrations on a sample of local analogs of high-redshift galaxies (for which auroral lines are detected in stacked spectra), selected to match the location of observed $z\sim 2$ sources in the \niibpt diagram \citep{shapley_mosdef_2015,strom_nebular_2017}; such calibrations, though still based on local galaxies, are thus specifically built for being applied to high-redshift galaxy spectra.

The two $z\sim7.6$ sources, ID$10612$ and ID$6533$, present a higher level of nebular excitation than lower redshift calibration samples, placing them above the upper plateau of R23, R3 and O32 curves, and below the expected R2 value at given metallicity, in the \cite{curti_massmetallicity_2020}, \cite{maiolino_amaze_2008}, \cite{nakajima_empress_2022}, and \cite{sanders_mosdef_mzr_2021} calibrations, whereas the \cite{bian_ldquodirectrdquo_2018} curves better cover that region of the line-ratios parameter space.
The extremely metal-poor $z\sim8.5$ ID$4590$ source, instead, is located in a region outside the validity range of most calibrations under study; 
nonetheless, it exhibits line-ratios properties resembling those of extremely metal-poor galaxies in the local Universe, and especially \cite{maiolino_amaze_2008} and \cite{nakajima_empress_2022} calibrations, as well as some of the extrapolations of \citealt{curti_massmetallicity_2020} and \citealt{sanders_mosdef_mzr_2021} curves (shown as dashed lines\footnote{It should be stressed that sometimes high-order polynomials express non-physical behaviours outside their fitting range}), provide a reasonable match to most of the line ratios observed in this galaxy.

Some of the observed deviations are actually still within the scatter of the calibration relations (typically ranging between $0.1$-$0.2$~dex in a given line ratio at fixed metallicity, depending on the diagnostic and the calibration sample). 
In Table~\ref{table:calibrations_offset}, we compare the observed and predicted (given the measured metallicity) line ratios for each considered strong-line diagnostics (i.e., the vertical offset from the calibration relation), and report the significance of the deviation (in units of $\sigma$) taking into account both measurements uncertainties and the intrinsic dispersion of each individual calibration\footnote{Since \citealt{bian_ldquodirectrdquo_2018} do not provide an estimate of the dispersion of the calibrations, we assume a typical dispersion of $0.15$~dex for all diagnostics. We further note that in some cases, e.g., the Ne$_{3}$O$_{2}$ calibration from \citealt{nakajima_empress_2022}, a very large intrinsic dispersion of the calibration may artificially lower the significance of the computed deviation, which is therefore not a good representation of the calibration accuracy.}.
In many cases, the measurements clearly fail to simultaneously match all the relations, prompting for a self-consistent revision of the calibrations at such early epochs which could be applied to galaxy spectra at different redshifts and/or observed in different filters, hence probing different emission lines.

\begin{table}
\caption{Significance of the deviation (in units of $\sigma$) of the observed line ratios in the \textit{JWST} z$\sim$8 galaxy sample from the predictions of each of the strong-line calibrations presented in Fig.~\ref{fig:calibrations}. Both the measurements uncertainties on the line ratios and the intrinsic dispersion of the calibrations ($\sigma_{cal}$, reported here as provided by the various authors) are taken into account.}
\centering
\label{table:calibrations_offset}
\begin{tabular}{@{}|l|ccccc@{}}
\hline\hline
Galaxy ID & R2 & R3 & O32 & R23 & Ne3O2 \\
\hline\hline

& \multicolumn{5}{|c|}{\citealt{maiolino_amaze_2008}} 
\vspace{0.05cm}\\
$\sigma_{cal}$ & 0.10 & 0.10 & 0.20 & 0.06 & 0.06 \vspace{0.1cm}\\
\hline

4590 & 1.68 &  3.03 &  0.08 &  1.77 &   0.39 \\
6355 &  2.31 &  1.88 &  4.39 &  0.96 &   3.16 \\
10612 &  2.20 &  0.86 &  2.97 &  0.41 &   3.18 \\
\hline\hline

& \multicolumn{5}{|c|}{\citealt{curti_new_2017, curti_massmetallicity_2020}} \vspace{0.05cm}\\

$\sigma_{cal}$ & 0.11 & 0.09 & 0.15 & 0.06 & 0.12 \vspace{0.1cm}\\
\hline
4590 & 14.94 &  3.22 &  2.54 &  0.59 & 1.97 \\
6355 & 2.99 &  2.62 &  3.80 &  1.80 & 3.82 \\
10612 & 1.89 &  1.80 &  2.11 &  0.64 & 2.04 \\
\hline
\hline
& \multicolumn{5}{|c|}{\citealt{bian_ldquodirectrdquo_2018}} 
\vspace{0.05cm}\\
$\sigma_{cal}$$^{1}$ & -- & -- & -- & -- & -- \vspace{0.1cm}\\
\hline

4590 & -- & 0.35 &  5.23 &  1.34 &   4.7 \\
6355 & -- & 1.50 &  3.71 &  0.95 &   2.52 \\
10612 & -- & 0.01 &  0.57 &  0.004 &   0.91 \\ 
\hline\hline
& \multicolumn{5}{|c|}{\citealt{sanders_mosdef_mzr_2021}} 
\vspace{0.05cm}\\
$\sigma_{cal}$ & 0.13 & 0.10 & 0.19 & 0.08 & 0.20 \vspace{0.1cm}\\
\hline

4590 &  3.10 &  2.31 &  1.98 &  1.37 &   0.92 \\
6355 & 3.65 &  2.86 &  4.05 &  1.77 &   2.64 \\
10612 & 3.29 &  3.86 &  4.32 &  1.49 &   4.28 \\
\hline\hline
& \multicolumn{5}{|c|}{\citealt{nakajima_empress_2022}} 
\vspace{0.05cm}\\
$\sigma_{cal}$ & 0.27 & 0.16 & 0.39 & 0.10 & 0.42 \vspace{0.1cm}\\
\hline

4590 &  0.95 &  2.54 &  1.62 &  1.18 &   0.95 \\
6355 & 1.26 &  1.75 &  1.72 &  1.38 &   0.18 \\
10612 & 1.29 &  0.74 &  1.32 &  0.53 &   1.58 \\
\hline\hline

\end{tabular}
{

\raggedright 
$^{1}$ For \cite{bian_ldquodirectrdquo_2018} we assume a dispersion of $0.15$~dex on each calibration.
\par}
\end{table}

\section{Cosmic Evolution of the Metallicity Scaling Relations}
\label{section:Evolution_of_the_metallicity_scaling_relations}

\subsection{The Mass-metallicity relation}

In this section we investigate the evolution in the metallicity scaling relations of galaxies as probed by the \Te-based O/H measurements enabled by \textit{JWST}/NIRSpec at $z\sim8$. 
In Fig.~\ref{fig:mzr}, we plot the three galaxies analysed in the current paper on the mass-metallicity plane, along with the local relation inferred for the SDSS sample by \citealt{curti_massmetallicity_2020} from \Te-based calibrations (contours in grey, best-fit MZR in black).
Compared to the MZR in the local Universe probed by SDSS galaxies, galaxies at $z\sim8$ appear metal deficient at fixed stellar mass, in agreement with the expected redshift evolution of the MZR \citep[see][and references therein]{maiolino_re_2019}.
More specifically, ID$6355$ is offset by $0.18$ dex from the $z\sim0$ MZR, whereas ID$10612$ is offset by $0.52$ dex. 
We also compare these observations at $z\sim8$ with the MZR at $z\sim 2$ and $z\sim3.3$, here parametrised by a linear regression fit performed on stacked spectra from the MOSDEF survey as presented in \cite{sanders_mosdef_mzr_2021}. These authors adopt the \cite{bian_ldquodirectrdquo_2018} \Te-empirical calibrations based on local analogues of high-z galaxies in order to reduce the potential biases affecting the use of local strong-line metallicity diagnostics at high redshift.
Extrapolating the \cite{sanders_mosdef_mzr_2021} relation to the low-mass range, we find that the metallicity of 
ID$6355$ is broadly consistent with the $z\sim2.2$ MZR, 
whereas ID$10612$ is offset by $0.12$ dex from the $z\sim3.3$ MZR extrapolation.
Naively, this could be interpreted as a signature of little or no 
evolution of the low-mass end of the MZR (either in slope or normalisation) between $z\sim 3.3$ and $z\sim 7.6$, possibly implying a mild cosmic evolution in both the gas fraction (and consequent metal dilution) and the metal-loading factor of star-formation-driven outflows over $\sim 1.2$ Gyr \citep{sanders_mosdef_mzr_2021}.


In contrast, we find that galaxy ID$4560$ is more than 1~dex offset from the z$\sim$0 MZR, and deviates strongly ($\sim$0.8~dex) also from the extrapolation of the MZR at z$\sim$3.3. Rather than very rapid redshift evolution of the MZR, this could indicate that ID$4560$ is in a very early evolutionary stage, in the process of rapidly building up its metals and approaching the MZR within a timeframe of just a few tens Myr \citep[e.g. Fig. 11 in ][]{maiolino_amaze_2008}. Alternatively, this finding could indicate that the slope of the MZR becomes steeper at very low masses (not probed by the z$\sim$2.2-3.3 surveys).

\subsection{The fundamental metallicity relation}

In this section we explore whether these galaxies at $z\sim8$ are consistent with the Fundamental Metallicity Relation (FMR), which (as discussed in the introduction) describes the correlation between metallicity, stellar mass, and SFR, observational consequence of the interplay between gas accretion, star formation, and outflows which govern the secular evolution of galaxies, and which was found not to evolve (or marginally evolve), out to z$\sim$3.
Fig.~\ref{fig:fmr} shows the deviation of the measured galaxy metallicities from the predictions of the FMR, as parameterized by equation 5 of \citet{curti_massmetallicity_2020}.
Different samples with available rest-frame optical spectroscopy compiled from the literature at various redshifts are included in the attempt to trace the cosmic evolution of this scaling relation across almost the entire history of the Universe. 
Specifically, these galaxies are compiled from \citet[][zCOSMOS at z$\sim$0.3--0.6]{cresci_metallicity_2012}, \citet[][KLEVER at z$\sim$2.2]{hayden-pawson_NO_2022}, \citet[][MOSDEF at z$\sim$2.3--3.3]{sanders_mosdef_mzr_2021}, \citet[][z$\sim$3.3]{onodera_ism_2016}, and \citet[][AMAZE at z$\sim$3.3]{troncoso_metallicity_2014}.
The metallicities of galaxies at $z>1$ have been consistently re-computed adopting the \cite{bian_ldquodirectrdquo_2018} \Te-based strong-line calibrations (to account for potential evolution in the diagnostics), whereas we adopt the \cite{curti_massmetallicity_2020} calibrations for SDSS and zCOSMOS galaxies.
For comparison, adopting instead the \cite{curti_massmetallicity_2020} calibrations for all galaxy samples produces, on average, a $\sim0.1$~dex lower metallicity in z$\sim2-3$ galaxies.
The various symbols in Fig.~\ref{fig:fmr} mark, for each considered literature dataset, the average offset in O/H (and standard deviation as errorbars) from the FMR predictions for individual galaxies within the sample. 
Red symbols mark instead individual \Te-based measurements for the \textit{JWST} galaxy sample at $z\sim8$; the errorbars on such datapoints are obtained by co-adding in quadrature the uncertainty on the log(O/H) measurement from the \Te method and the uncertainty on the metallicity predicted by the FMR, evaluated from the standard deviation of the distribution of metallicities obtained by varying one hundred times the input \mstar\ and SFR of galaxies within their errors.

The two galaxies at $z\sim$~7.6 show offsets of about $\sim$0.2~dex (significant at $\sim$2$\sigma$) from the FMR, though with different signs, suggesting different evolutionary stages.
However, we also note that considering additional sources of uncertainty, like the errors associated to the parametrisation of the functional form of the FMR and its extrapolation to the average \mstar and SFR of the \textit{JWST} sample (i.e., $\sim0.25$~dex uncertainty, cfr. Fig.~11 in \citealt{curti_massmetallicity_2020}), or the systematics introduced by the existence of temperatures stratification biasing the \Te-derived metallicity \citep{kewley_understanding_2019}, would make the two z$\sim7.6$ galaxies broadly consistent with the FMR predictions.

The situation is readily different for the $z\sim8.5$ source ID$4590$, which is observed to be much more metal-poor than the other two. This galaxy falls at the very low-mass, low-metallicity end of the MZR in Fig.~\ref{fig:mzr} and is also offset by $\sim$0.9~dex from the metallicity expected by the FMR given its \mstar\ and SFR, a deviation significant at $\sim$6$\sigma$ ($>3\sigma$ even accounting for uncertainties on the FMR extrapolation).
Considering that only $\sim 100$\,Myr separate the two epochs (i.e., $z\sim7.6$ and $z\sim8.5$), this suggests that this object might be observed very far from the equilibrium between chemical enrichment and gas flows, 
and it is likely experiencing an initial, steeply rising phase of enrichment (see discussion in previous section) while being swamped by accretion of pristine gas.
In agreement with the emission-line properties already discussed in terms of the metallicity calibration plots, the properties of this object resemble those of extremely metal-poor galaxies observed in the local Universe, which indeed have been long considered as potential analogues of galaxies in the epoch of reionisation \citep[e.g.][]{Izotov2018, izotov_low-redshift_2019,izotov_xmp_2021}.


Overall, we have observed three galaxies at z$\sim$8 showing a large scatter in metallicity, probably as a consequence of their early evolutionary stage, questioning whether the FMR scaling relation observed at lower redshifts, and associated with a more smooth and secular evolution, is already in place at these cosmic epochs.
However, much larger and statistically robust samples are required in order to draw any strong conclusion on the evolution of such scaling relation at these redshifts.

\subsection{Comparison with theoretical predictions}

\begin{figure}
    \centering
  \includegraphics[width=\columnwidth]{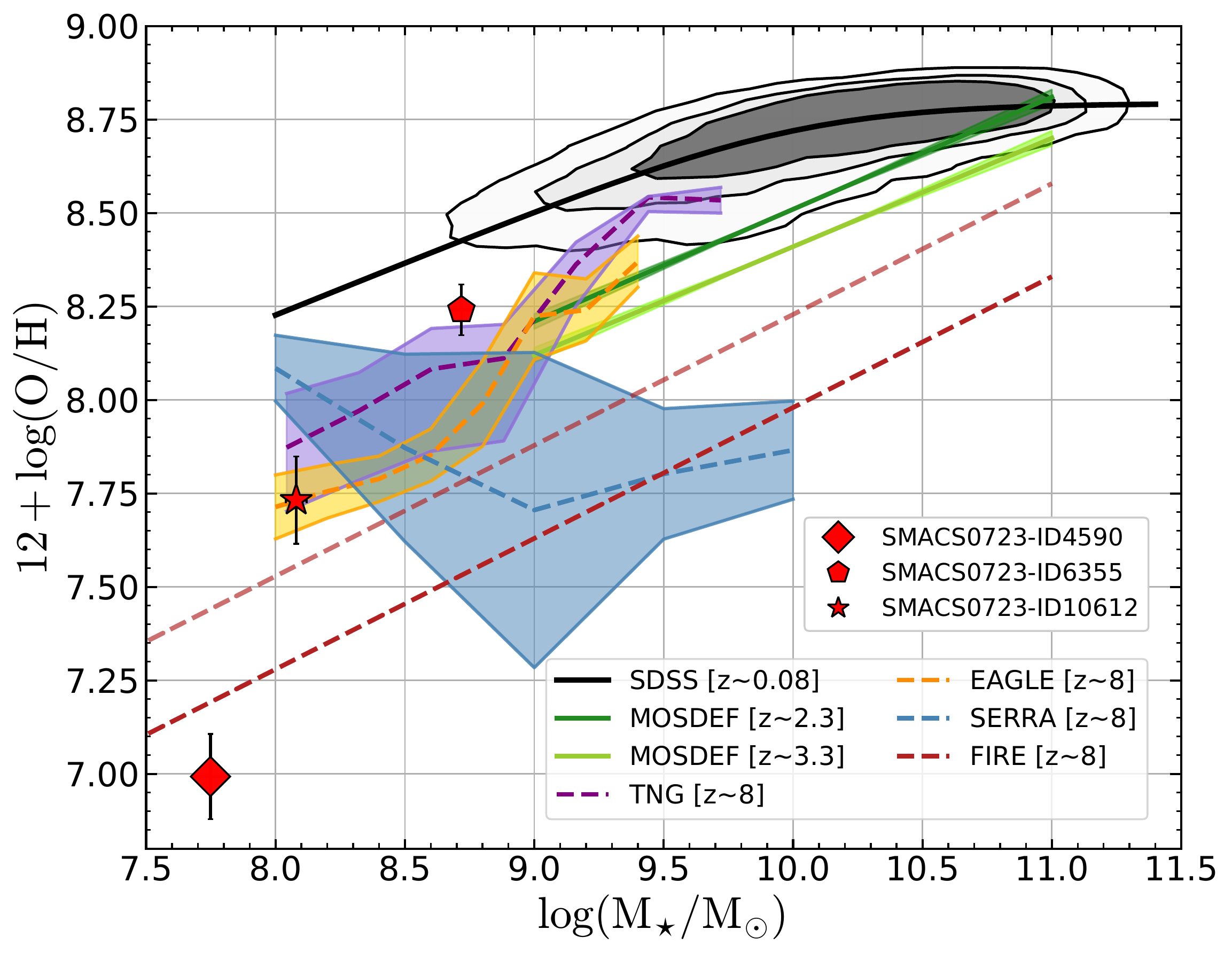}
    \caption{Mass-Metallicity relation (MZR) for the \textit{JWST} sample. 
    The MZR for SDSS galaxies from \citealt{curti_massmetallicity_2020} and its best-fit are shown by the grey contours and black curve, respectively. 
    The best-fit to stacked spectra of MOSDEF galaxies from \citealt{sanders_mosdef_mzr_2021} (based on the \citealt{bian_ldquodirectrdquo_2018} calibrations tuned for high-z) are shown as representative for the MZR at $z\sim2.3$ and $z\sim3.3$.
    Moreover, the MZR at $z\sim8$ predicted by TNG \citep{nelson_TNG_2019}, EAGLE \citep{Schaye_EAGLE_2015}, FIRE \citep{ma_origin_2016} (extrapolated to z=8), and SERRA (\citealt{pallottini_serra_2022}) simulations are shown as dashed purple, orange, brown (light brown for re-scaled relation to match both local and z=2--3 observations) and blue lines, respectively, with the shaded areas marking the region between the 16th and 84th percentiles of the predicted relation. 
    }
    \label{fig:mzr}
\end{figure}

\begin{figure}
    \centering
\includegraphics[width=\columnwidth]{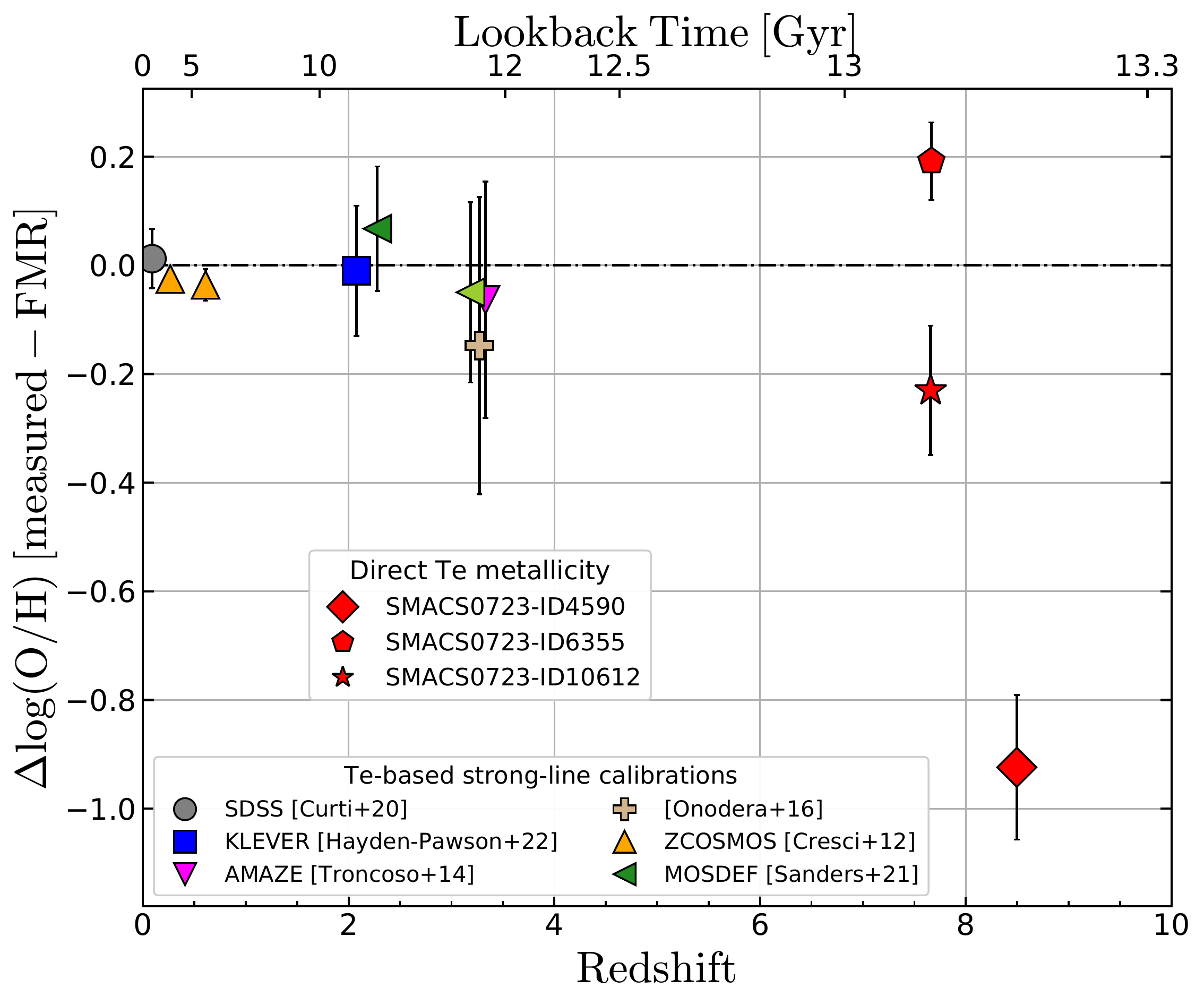}
    \caption{Deviations in the observed log(O/H) from the predictions of the local Fundamental Metallicity Relation (FMR, from \citealt{curti_massmetallicity_2020}, equation 5), plotted as a function of redshift (look back time is reported on the top axis). High-redshift galaxies with \Te-based measurements from \textit{JWST}/NIRSpec are shown as red symbols.
    The average (and standard deviation) offset computed for different datasets complied from the literature at different redshifts are also shown for comparison. All metallicities from the literature samples are self-consistently re-computed adopting the \Te-based \citealt{curti_massmetallicity_2020} calibrations at z$<1$ and the \citealt{bian_ldquodirectrdquo_2018} at z$>1$. At z$\sim8$, galaxies appear offset from the predictions of the local FMR (though with different levels of significance), suggesting that they are far from the smooth equilibrium between chemical enrichment and gas flows that sets the average scaling relations in local galaxies (and up to z$\sim3$).
    }
    \label{fig:fmr}
\end{figure}

Finally, in Fig.~\ref{fig:mzr} we also compare our observations with the predictions of the MZR at $z\sim8$ extracted from different suites of cosmological box and zoom-in simulations, namely IllustrisTNG \citep{Naiman2018, Nelson2018, nelson_TNG_2019, Pillepich2018, Springel2018, Marinacci2018}, EAGLE \citep{Crain2015, Schaye_EAGLE_2015, McAlpine2016}, FIRE \citep{ma_origin_2016}, and SERRA \citep{pallottini_serra_2022}. 
In both EAGLE and IllustrisTNG we make use of the publicly available subhalo catalogues at $z=8$ from the highest resolution runs in $\sim 100^3\ \rm cMpc^3$ cosmological boxes with fiducial subgrid physics prescriptions. 
In order to calculate gas phase metallicities in IllustrisTNG and EAGLE, we adopt the approach of \cite{Torrey2019}, considering central galaxies and assuming that oxygen comprises 35\% of the SFR-weighted metal mass fraction within twice the stellar half-mass radius (IllustrisTNG) and 35\% of the metal mass fraction in star-forming gas particles in a bound halo (EAGLE).
FIRE simulations instead define the gas-phase metallicity as the mass-weighted metallicity of all gas particles that belong to the ISM (defined by a temperature below $10^{4}$~K, \citealt{ma_origin_2016}, whereas in SERRA the metallicity is tracked as the sum of all heavy elements and gas, and stellar metallicity are coupled \citep{pallottini_serra_2022}.
Both FIRE and SERRA assume solar abundance ratios \citep{asplund_solar_2009}.

In Fig.~\ref{fig:mzr} we find that both EAGLE and IllustrisTNG are in reasonably good agreement with the observations of the two $z\sim7.6$ galaxies with \Te measurements. 
In order to contextualise this result on the absolute metallicity scale, we have verified that our physical assumptions on oxygen yield provides good matches to both the parametrisation of the MZR of \citet{curti_massmetallicity_2020} at $z\sim0.08$ and of \cite{sanders_mosdef_mzr_2021} at $z=2$--$3$, which are based on \Te-calibrated strong-line diagnostics.
This comparison hence suggests that both suites may capture some of the primary chemical enrichment in place in the early Universe, despite the lack of observational data available for model calibration at high redshift.

The redshift-evolution of the MZR predicted by FIRE simulations \citep{ma_origin_2016} (originally tracing galaxies from $z=0$--$6$ and here extrapolated to z$\sim$8) shows instead a much lower normalisation, suggesting a strong evolution between $z\sim3.3$ and $z\sim8$, and getting closer to what we observe in ID$4590$.
However, in contrast to what is found for EAGLE and IllustrisTNG, we find that the FIRE-based MZR parametrisation from \cite{ma_origin_2016} predicts $0.22$~dex and $0.24$~dex lower metallicities than observed in MOSDEF galaxies at $z\sim2.3$ and $z\sim3.3$ from \cite{sanders_mosdef_mzr_2021}, respectively, whereas $0.15$~dex higher normalisation compared to the SDSS-MZR at $z\sim0.08$ from \cite{curti_massmetallicity_2020}. We attempt to correct for such redshift-dependent offset by fitting an exponential function of the form [A $e^{-z/z_{0}}$ + C]\footnote{best-fit parameters: A$=0.43$; z$_{0}$=0.85; C$=-0.25$} to the measured $\upDelta$log(O/H) in FIRE simulations at the three redshifts ($z\sim0.08$, $z\sim2.3$, $z\sim3.3$), and use it to predict a normalisation offset at $z\sim8$ of $\upDelta$log(O/H)$=-0.25$~dex. In Fig.\ref{fig:mzr}, we also show such re-scaled z$\sim$8 extrapolation of the FIRE-MZR as the light-brown dashed line, which gets closer to what observed in ID10612.
In contrast, the high-resolution ($25$\,pc at $z=8$) SERRA suite of cosmological zoom-in simulations predict a flatter MZR at $z\sim8$ with much larger dispersion, suggesting such a scaling relation is not yet in place at these cosmic epochs. We note that the SERRA simulations follow galaxies only down to $z\sim6$, hence no direct comparison with lower redshift observations can be performed.
Needless to say, more metallicity determinations at such early epochs and more dedicated simulations of high-redshift galaxy populations are needed to constrain the theoretical predictions on the evolution of the MZR at very high-z relative to its shape at lower redshift.

\section{Conclusions}
\label{sec:conclusions}
We have analysed the properties of the three $z\sim$~8 gravitationally lensed galaxies observed in the framework of the \textit{JWST} Early Release Observations. These galaxies were selected based solely on their redshift z$>$7.
We have used sensitive NIRSpec spectra probing the rest-frame optical emission in these sources.
 Through a careful processing and inspection of the spectra, we could study the ISM properties in these galaxies, and apply the robust \Te-method to infer the gas-phase oxygen abundance for the first time at such high redshift.
Our are main findings for these three galaxies at z$\sim$8 are summarised as follows:

\begin{itemize}
    \item The Balmer lines are consistent with some, low level of dust extinction (Fig.~\ref{fig:balmer_decrements}). This finding is also consistent with the analysis of the continuum SED.
    \item The excitation diagnostic ratios, such as [\ion{O}{iii}]/H$\beta$, [\ion{O}{ii}]/H$\beta$ and [\ion{O}{iii}]/[\ion{O}{ii}] are in the tail of the distribution observed in galaxies at z$\sim$2--3 (Fig.~\ref{fig:R3R2}). Remarkably, some of them have very low [\ion{O}{ii}]/H$\beta$ and very high [\ion{O}{iii}]]/[[\ion{O}{ii}]. These are consistent with the expectation of photoionisation models for metal-poor galaxies and/or galaxies with high escape fraction of ionising, Lyman continuum photons.
    \item We robustly detect the [\ion{O}{iii}]$\lambda$4363 auroral line in all three galaxies, with ratios relative to the strong [\ion{O}{iii}]$\lambda\lambda$4959,5007 doublet that are large relative to local galaxies (0.01--0.04).
    \item The inferred gas temperatures range from 1.2~10$^{4}$~K to 2.8~10$^{4}$~K, with the \Te-based metallicities ranging from extremely metal poor (12+log(O/H)$\sim$7) to about one third solar.
    \item None of the local strong-line metallicity calibrations seem to provide a good prediction of the observed metallicities at z$\sim$8 simultaneously across all diagnostics and for all galaxies (Fig.~\ref{fig:calibrations}). Different calibrations seem to perform better for some galaxies or for some diagnostic ratios, suggesting that a systemic re-calibration of the strong-line diagnostics is needed for these early epochs.
    \item The two most ``massive'' galaxies of the sample at z$\sim$7.6 (log(\mstar/M$_{\odot}$)=$8.1$-$8.7$) are scattered around the mass-metallicity relation at z$\sim$2-3 (Fig.~\ref{fig:mzr}), potentially suggestive of little evolution of the MZR from z$\sim$2--3 to z$\sim$8. On the contrary, the least massive and most metal-poor galaxy (log(\mstar/M$_{\odot}$)=$7.7$, 12+log(O/H)=$6.99$) deviates significantly from the relation at lower redshifts, suggesting that this galaxy is in a phase of rapid evolution or that the MZR has a much steeper slope at such low masses.
    Different cosmological simulations can reproduce some of the galaxies on the MZR, but none of them seem capable of simultaneously matching all of them.
    \item The three galaxies are widely scattered relative to the Fundamental Metallicity Relation, with two of them marginally consistent and one (ID$4590$ at z$\sim8.5$) strongly deviating (Fig.~\ref{fig:fmr}). This finding suggest that galaxies at such early epochs are undergoing rapidly evolving mechanisms, and not yet settled on the local scaling relations that are instead associated with smooth, secular evolutionary processes. 
    In any case, much larger samples are required in order to draw strong conclusions on the evolution of such scaling relations.
\end{itemize}

In conclusion, these first datasets already highlight the giant leap in the redshift parameter space brought to the field by the advent of rest-frame optical spectroscopy enabled by \textit{JWST}, which is about to open a window on new modes of galaxy formation and evolution, helping to discover objects with properties radically different than all the other known galaxies. 
Several forthcoming observational programmes will target the `redshift desert' between z$\sim$3--8 (and beyond), allowing us to obtain a complete and coherent picture of the evolution of the chemical properties of galaxies across the entire cosmic history.

Nonetheless, we also stress that assessing the applicability of the strong-line metallicity diagnostics at high redshift remains an open problem, which requires both more a carefully selected sample of galaxies and deeper observations to detect auroral lines over a wider range of metallicities.
While the observations discussed in the present paper provide a preview of the possibilities opened by \textit{JWST} in this sense, a variety of programmes scheduled in Cycle 1 primarily designed to address this science topic \citep[e.g.][]{2021jwst.prop.1879C,2021jwst.prop.2593S,2021jwst.prop.1914S} will soon provide a more robust re-calibration of the strong-line metallicity diagnostics for the high-redshift Universe.

\section*{Acknowledgements}
We thank the anonymous referee for his/her comments which contributed to improving this paper.

M.C., R.M., F.D'E., L.S., J.S., W.B., J.W., T.J.L., C.H-P. and J.S.B. acknowledge support by the Science and Technology Facilities Council (STFC) and ERC Advanced Grant 695671 "QUENCH". RM also acknowledges funding from a research professorship from the Royal Society.
S.C.  acknowledges support from
the ERC Advanced Grant INTERSTELLAR H2020/740120.
H\"U gratefully acknowledges support by the Isaac Newton Trust and by the Kavli Foundation through a Newton-Kavli Junior Fellowship.
SA acknowledges funding from grant PID2021-127718NB-I00 by the Spanish Ministry of Science and Innovation/State Agency of Research (MCIN/AEI).
J.M.P and J.W. gratefully acknowledge support from the MERAC Foundation. AJB acknowledges funding from the ``FirstGalaxies" Advanced Grant from the European Research Council (ERC) under the European Union’s Horizon 2020 research and innovation programme (Grant agreement No. 789056). RS acknowledges support from an STFC Ernest Rutherford Fellowship (ST/S004831/1).
ECL acknowledges support of an STFC Webb Fellowship (ST/W001438/1).
We are grateful to Pierre Ferruit, Peter Jakobsen, and Nora L\"{u}tzgendorf for sharing their expertise on NIRSpec and the processing of its unique data.

This work is based on observations made with the NASA/ESA/CSA \textit{James Webb Space Telescope}.
Some of the data presented in this paper were obtained from the Multimission Archive at the Space Telescope Science Institute (MAST). STScI is operated by the Association of Universities for Research in Astronomy, Inc., under NASA contract NAS5-26555. Support for MAST for non-\textit{HST} data is provided by the NASA Office of Space Science via grant NAG5-7584 and by other grants and contracts.

Finally, M.C. is heartily grateful to Vittoria for showing him a completely new and wonderful, yet unknown and mysterious, part of the Universe.

\section*{Data Availability}

The raw (level 2) data underlying this article are publicly available from the \hyperlink{https://mast.stsci.edu/portal/Mashup/Clients/Mast/Portal.html}{MAST} archive.
The processed and calibrated 1D spectra analysed in this work are publicly available at \url{https://doi.org/10.5281/zenodo.6940561}. 
Other data are available upon reasonable request to the corresponding author.



\bibliographystyle{mnras}
\bibliography{main} 

\begin{thebibliography}{}
\makeatletter
\relax
\def\mn@urlcharsother{\let\do\@makeother \do\$\do\&\do\#\do\^\do\_\do\%\do\~}
\def\mn@doi{\begingroup\mn@urlcharsother \@ifnextchar [ {\mn@doi@}
  {\mn@doi@[]}}
\def\mn@doi@[#1]#2{\def\@tempa{#1}\ifx\@tempa\@empty \href
  {http://dx.doi.org/#2} {doi:#2}\else \href {http://dx.doi.org/#2} {#1}\fi
  \endgroup}
\def\mn@eprint#1#2{\mn@eprint@#1:#2::\@nil}
\def\mn@eprint@arXiv#1{\href {http://arxiv.org/abs/#1} {{\tt arXiv:#1}}}
\def\mn@eprint@dblp#1{\href {http://dblp.uni-trier.de/rec/bibtex/#1.xml}
  {dblp:#1}}
\def\mn@eprint@#1:#2:#3:#4\@nil{\def\@tempa {#1}\def\@tempb {#2}\def\@tempc
  {#3}\ifx \@tempc \@empty \let \@tempc \@tempb \let \@tempb \@tempa \fi \ifx
  \@tempb \@empty \def\@tempb {arXiv}\fi \@ifundefined
  {mn@eprint@\@tempb}{\@tempb:\@tempc}{\expandafter \expandafter \csname
  mn@eprint@\@tempb\endcsname \expandafter{\@tempc}}}

\bibitem[\protect\citeauthoryear{Allende~Prieto, Lambert  \&
  Asplund}{Allende~Prieto et~al.}{2001}]{allende_prieto_forbidden_2001}
Allende~Prieto C.,  Lambert D.~L.,   Asplund M.,  2001, \mn@doi [\apjl]
  {10.1086/322874}, 556, L63

\bibitem[\protect\citeauthoryear{Andrews \& Martini}{Andrews \&
  Martini}{2013}]{andrews_mass-metallicity_2013}
Andrews B.~H.,  Martini P.,  2013, \mn@doi [\apj]
  {10.1088/0004-637X/765/2/140}, 765, 140

\bibitem[\protect\citeauthoryear{{Asplund}, {Grevesse}, {Sauval}  \&
  {Scott}}{{Asplund} et~al.}{2009}]{asplund_solar_2009}
{Asplund} M.,  {Grevesse} N.,  {Sauval} A.~J.,   {Scott} P.,  2009, \mn@doi
  [\araa] {10.1146/annurev.astro.46.060407.145222}, \href
  {https://ui.adsabs.harvard.edu/abs/2009ARA&A..47..481A} {47, 481}

\bibitem[\protect\citeauthoryear{Baldwin, Phillips  \& Terlevich}{Baldwin
  et~al.}{1981}]{baldwin_classification_1981}
Baldwin J.~A.,  Phillips M.~M.,   Terlevich R.,  1981, \mn@doi [\pasp]
  {10.1086/130766}, 93, 5

\bibitem[\protect\citeauthoryear{{Barrow}, {Robertson}, {Ellis}, {Nakajima},
  {Saxena}, {Stark}  \& {Tang}}{{Barrow} et~al.}{2020}]{Barrow2020}
{Barrow} K. S.~S.,  {Robertson} B.~E.,  {Ellis} R.~S.,  {Nakajima} K.,
  {Saxena} A.,  {Stark} D.~P.,   {Tang} M.,  2020, \mn@doi [\apjl]
  {10.3847/2041-8213/abbd8e}, \href
  {https://ui.adsabs.harvard.edu/abs/2020ApJ...902L..39B} {902, L39}

\bibitem[\protect\citeauthoryear{Berg et~al.,}{Berg
  et~al.}{2012}]{berg_direct_2012}
Berg D.~A.,  et~al., 2012, \mn@doi [\apj] {10.1088/0004-637X/754/2/98}, 754, 98

\bibitem[\protect\citeauthoryear{Bian, Kewley  \& Dopita}{Bian
  et~al.}{2018}]{bian_ldquodirectrdquo_2018}
Bian F.,  Kewley L.~J.,   Dopita M.~A.,  2018, \mn@doi [\apj]
  {10.3847/1538-4357/aabd74}, 859, 175

\bibitem[\protect\citeauthoryear{{Bohlin}, {Gordon}  \& {Tremblay}}{{Bohlin}
  et~al.}{2014}]{Bohlin14}
{Bohlin} R.~C.,  {Gordon} K.~D.,   {Tremblay} P.~E.,  2014, \mn@doi [\pasp]
  {10.1086/677655}, \href
  {https://ui.adsabs.harvard.edu/abs/2014PASP..126..711B} {126, 711}

\bibitem[\protect\citeauthoryear{{Bohlin}, {Hubeny}  \& {Rauch}}{{Bohlin}
  et~al.}{2020}]{Bohlin20}
{Bohlin} R.~C.,  {Hubeny} I.,   {Rauch} T.,  2020, \mn@doi [\aj]
  {10.3847/1538-3881/ab94b4}, \href
  {https://ui.adsabs.harvard.edu/abs/2020AJ....160...21B} {160, 21}

\bibitem[\protect\citeauthoryear{{Bouch{\'e}} et~al.,}{{Bouch{\'e}}
  et~al.}{2010}]{bouche_cold_accretion_2010}
{Bouch{\'e}} N.,  et~al., 2010, \mn@doi [\apj] {10.1088/0004-637X/718/2/1001},
  \href {https://ui.adsabs.harvard.edu/abs/2010ApJ...718.1001B} {718, 1001}

\bibitem[\protect\citeauthoryear{Brinchmann, Charlot, White, Tremonti,
  Kauffmann, Heckman  \& Brinkmann}{Brinchmann
  et~al.}{2004}]{brinchmann_physical_2004}
Brinchmann J.,  Charlot S.,  White S. D.~M.,  Tremonti C.,  Kauffmann G.,
  Heckman T.,   Brinkmann J.,  2004, \mn@doi [\mnras]
  {10.1111/j.1365-2966.2004.07881.x}, 351, 1151

\bibitem[\protect\citeauthoryear{{Caminha}, {Suyu}, {Mercurio}, {Brammer},
  {Bergamini}, {Vanzella}  \& {Acebron}}{{Caminha} et~al.}{2022}]{Caminha2022}
{Caminha} G.~B.,  {Suyu} S.~H.,  {Mercurio} A.,  {Brammer} G.,  {Bergamini} P.,
   {Vanzella} E.,   {Acebron} A.,  2022, arXiv e-prints, \href
  {https://ui.adsabs.harvard.edu/abs/2022arXiv220707567C} {p. arXiv:2207.07567}

\bibitem[\protect\citeauthoryear{Cappellari}{Cappellari}{2017}]{cappellari_improving_2017}
Cappellari M.,  2017, \mn@doi [\mnras] {10.1093/mnras/stw3020}, 466, 798

\bibitem[\protect\citeauthoryear{{Carnall}, {McLure}, {Dunlop}  \&
  {Dav{\'e}}}{{Carnall} et~al.}{2018}]{2018MNRAS.480.4379C}
{Carnall} A.~C.,  {McLure} R.~J.,  {Dunlop} J.~S.,   {Dav{\'e}} R.,  2018,
  \mn@doi [\mnras] {10.1093/mnras/sty2169}, \href
  {https://ui.adsabs.harvard.edu/abs/2018MNRAS.480.4379C} {480, 4379}

\bibitem[\protect\citeauthoryear{{Carnall} et~al.,}{{Carnall}
  et~al.}{2022}]{2022arXiv220708778C}
{Carnall} A.~C.,  et~al., 2022, arXiv e-prints, \href
  {https://ui.adsabs.harvard.edu/abs/2022arXiv220708778C} {p. arXiv:2207.08778}

\bibitem[\protect\citeauthoryear{{Chabrier}}{{Chabrier}}{2003}]{2003PASP..115..763C}
{Chabrier} G.,  2003, \mn@doi [\pasp] {10.1086/376392}, \href
  {https://ui.adsabs.harvard.edu/abs/2003PASP..115..763C} {115, 763}

\bibitem[\protect\citeauthoryear{{Chevallard} \& {Charlot}}{{Chevallard} \&
  {Charlot}}{2016}]{2016MNRAS.462.1415C}
{Chevallard} J.,  {Charlot} S.,  2016, \mn@doi [\mnras]
  {10.1093/mnras/stw1756}, \href
  {https://ui.adsabs.harvard.edu/abs/2016MNRAS.462.1415C} {462, 1415}

\bibitem[\protect\citeauthoryear{{Chisholm} et~al.,}{{Chisholm}
  et~al.}{2022}]{Chisholm_escape_frac_2022}
{Chisholm} J.,  et~al., 2022, arXiv e-prints, \href
  {https://ui.adsabs.harvard.edu/abs/2022arXiv220705771C} {p. arXiv:2207.05771}

\bibitem[\protect\citeauthoryear{{Choi}, {Dotter}, {Conroy}, {Cantiello},
  {Paxton}  \& {Johnson}}{{Choi} et~al.}{2016}]{choi_MIST_2016}
{Choi} J.,  {Dotter} A.,  {Conroy} C.,  {Cantiello} M.,  {Paxton} B.,
  {Johnson} B.~D.,  2016, \mn@doi [\apj] {10.3847/0004-637X/823/2/102}, \href
  {https://ui.adsabs.harvard.edu/abs/2016ApJ...823..102C} {823, 102}

\bibitem[\protect\citeauthoryear{Christensen et~al.,}{Christensen
  et~al.}{2012}]{christensen_gravitationally_2012}
Christensen L.,  et~al., 2012, \mn@doi [\mnras]
  {10.1111/j.1365-2966.2012.22007.x}, 427, 1973

\bibitem[\protect\citeauthoryear{{Conroy}, {Naidu}, {Zaritsky}, {Bonaca},
  {Cargile}, {Johnson}  \& {Caldwell}}{{Conroy}
  et~al.}{2019}]{conroy_stellar_halo_2019}
{Conroy} C.,  {Naidu} R.~P.,  {Zaritsky} D.,  {Bonaca} A.,  {Cargile} P.,
  {Johnson} B.~D.,   {Caldwell} N.,  2019, \mn@doi [\apj]
  {10.3847/1538-4357/ab5710}, \href
  {https://ui.adsabs.harvard.edu/abs/2019ApJ...887..237C} {887, 237}

\bibitem[\protect\citeauthoryear{{Crain} et~al.,}{{Crain}
  et~al.}{2015}]{Crain2015}
{Crain} R.~A.,  et~al., 2015, \mn@doi [\mnras] {10.1093/mnras/stv725}, \href
  {https://ui.adsabs.harvard.edu/abs/2015MNRAS.450.1937C} {450, 1937}

\bibitem[\protect\citeauthoryear{Cresci, Mannucci, Sommariva, Maiolino, Marconi
   \& Brusa}{Cresci et~al.}{2012}]{cresci_metallicity_2012}
Cresci G.,  Mannucci F.,  Sommariva V.,  Maiolino R.,  Marconi A.,   Brusa M.,
  2012, \mn@doi [\mnras] {10.1111/j.1365-2966.2011.20299.x}, 421, 262

\bibitem[\protect\citeauthoryear{{Cresci}, {Mannucci}  \& {Curti}}{{Cresci}
  et~al.}{2019}]{Cresci2019}
{Cresci} G.,  {Mannucci} F.,   {Curti} M.,  2019, \mn@doi [\aap]
  {10.1051/0004-6361/201834637}, \href
  {https://ui.adsabs.harvard.edu/abs/2019A&A...627A..42C} {627, A42}

\bibitem[\protect\citeauthoryear{Curti, Cresci, Mannucci, Marconi, Maiolino  \&
  Esposito}{Curti et~al.}{2017}]{curti_new_2017}
Curti M.,  Cresci G.,  Mannucci F.,  Marconi A.,  Maiolino R.,   Esposito S.,
  2017, \mn@doi [\mnras] {10.1093/mnras/stw2766}, 465, 1384

\bibitem[\protect\citeauthoryear{Curti, Mannucci, Cresci  \& Maiolino}{Curti
  et~al.}{2020a}]{curti_massmetallicity_2020}
Curti M.,  Mannucci F.,  Cresci G.,   Maiolino R.,  2020a, \mn@doi [Monthly
  Notices of the Royal Astronomical Society] {10.1093/mnras/stz2910}, 491, 944

\bibitem[\protect\citeauthoryear{Curti et~al.,}{Curti
  et~al.}{2020b}]{curti_klever_2020}
Curti M.,  et~al., 2020b, \mn@doi [Monthly Notices of the Royal Astronomical
  Society] {10.1093/mnras/stz3379}, 492, 821

\bibitem[\protect\citeauthoryear{{Curti} et~al.,}{{Curti}
  et~al.}{2021}]{2021jwst.prop.1879C}
{Curti} M.,  et~al., 2021, {Opening the era of direct metallicity measurements
  in high redshift galaxies}, JWST Proposal. Cycle 1, ID. \#1879

\bibitem[\protect\citeauthoryear{{Curti} et~al.,}{{Curti}
  et~al.}{2022}]{Curti22}
{Curti} M.,  et~al., 2022, \mn@doi [\mnras] {10.1093/mnras/stac544}, \href
  {https://ui.adsabs.harvard.edu/abs/2022MNRAS.512.4136C} {512, 4136}

\bibitem[\protect\citeauthoryear{Davé, Rafieferantsoa, Thompson  \&
  Hopkins}{Davé et~al.}{2017}]{dave_mufasa_2017}
Davé R.,  Rafieferantsoa M.~H.,  Thompson R.~J.,   Hopkins P.~F.,  2017,
  \mn@doi [\mnras] {10.1093/mnras/stx108}, 467, 115

\bibitem[\protect\citeauthoryear{{D{\'\i}az}, {Castellanos}, {Terlevich}  \&
  {Luisa Garc{\'\i}a-Vargas}}{{D{\'\i}az} et~al.}{2000}]{diaz_hii_regions_2000}
{D{\'\i}az} A.~I.,  {Castellanos} M.,  {Terlevich} E.,   {Luisa
  Garc{\'\i}a-Vargas} M.,  2000, \mn@doi [\mnras]
  {10.1046/j.1365-8711.2000.03737.x}, \href
  {https://ui.adsabs.harvard.edu/abs/2000MNRAS.318..462D} {318, 462}

\bibitem[\protect\citeauthoryear{{Dors}, {Contini}, {Riffel},
  {P{\'e}rez-Montero}, {Krabbe}, {Cardaci}  \& {H{\"a}gele}}{{Dors}
  et~al.}{2021}]{Dors2020}
{Dors} O.~L.,  {Contini} M.,  {Riffel} R.~A.,  {P{\'e}rez-Montero} E.,
  {Krabbe} A.~C.,  {Cardaci} M.~V.,   {H{\"a}gele} G.~F.,  2021, \mn@doi
  [\mnras] {10.1093/mnras/staa3707}, \href
  {https://ui.adsabs.harvard.edu/abs/2021MNRAS.501.1370D} {501, 1370}

\bibitem[\protect\citeauthoryear{{Ebeling}, {Edge}  \& {Henry}}{{Ebeling}
  et~al.}{2001}]{Ebeling2001}
{Ebeling} H.,  {Edge} A.~C.,   {Henry} J.~P.,  2001, \mn@doi [\apj]
  {10.1086/320958}, \href
  {https://ui.adsabs.harvard.edu/abs/2001ApJ...553..668E} {553, 668}

\bibitem[\protect\citeauthoryear{{Ebeling}, {Barrett}, {Donovan}, {Ma}, {Edge}
  \& {van Speybroeck}}{{Ebeling} et~al.}{2007}]{Ebeling2007}
{Ebeling} H.,  {Barrett} E.,  {Donovan} D.,  {Ma} C.~J.,  {Edge} A.~C.,   {van
  Speybroeck} L.,  2007, \mn@doi [\apjl] {10.1086/518603}, \href
  {https://ui.adsabs.harvard.edu/abs/2007ApJ...661L..33E} {661, L33}

\bibitem[\protect\citeauthoryear{{Ebeling}, {Edge}, {Mantz}, {Barrett},
  {Henry}, {Ma}  \& {van Speybroeck}}{{Ebeling} et~al.}{2010}]{Ebeling2010}
{Ebeling} H.,  {Edge} A.~C.,  {Mantz} A.,  {Barrett} E.,  {Henry} J.~P.,  {Ma}
  C.~J.,   {van Speybroeck} L.,  2010, \mn@doi [\mnras]
  {10.1111/j.1365-2966.2010.16920.x}, \href
  {https://ui.adsabs.harvard.edu/abs/2010MNRAS.407...83E} {407, 83}

\bibitem[\protect\citeauthoryear{{Ebeling} et~al.,}{{Ebeling}
  et~al.}{2013}]{Ebeling2013}
{Ebeling} H.,  et~al., 2013, \mn@doi [\mnras] {10.1093/mnras/stt387}, \href
  {https://ui.adsabs.harvard.edu/abs/2013MNRAS.432...62E} {432, 62}

\bibitem[\protect\citeauthoryear{Ellison, Patton, Simard  \&
  McConnachie}{Ellison et~al.}{2008}]{ellison_clues_2008}
Ellison S.~L.,  Patton D.~R.,  Simard L.,   McConnachie A.~W.,  2008, \mn@doi
  [\apjl] {10.1086/527296}, 672, L107

\bibitem[\protect\citeauthoryear{Erb, Pettini, Steidel, Strom, Rudie, Trainor,
  Shapley  \& Reddy}{Erb et~al.}{2016}]{erb_high_2016}
Erb D.~K.,  Pettini M.,  Steidel C.~C.,  Strom A.~L.,  Rudie G.~C.,  Trainor
  R.~F.,  Shapley A.~E.,   Reddy N.~A.,  2016, \mn@doi [\apj]
  {10.3847/0004-637X/830/1/52}, 830, 52

\bibitem[\protect\citeauthoryear{{Ferruit} et~al.,}{{Ferruit}
  et~al.}{2022}]{Ferruit2022}
{Ferruit} P.,  et~al., 2022, \mn@doi [\aap] {10.1051/0004-6361/202142673},
  \href {https://ui.adsabs.harvard.edu/abs/2022A&A...661A..81F} {661, A81}

\bibitem[\protect\citeauthoryear{Garnett}{Garnett}{1990}]{garnett_nitrogen_1990}
Garnett D.~R.,  1990, \mn@doi [\apj] {10.1086/169324}, 363, 142

\bibitem[\protect\citeauthoryear{{Gordon}, {Clayton}, {Misselt}, {Landolt}  \&
  {Wolff}}{{Gordon} et~al.}{2003}]{gordon_LMC_attenuation_2003}
{Gordon} K.~D.,  {Clayton} G.~C.,  {Misselt} K.~A.,  {Landolt} A.~U.,   {Wolff}
  M.~J.,  2003, \mn@doi [\apj] {10.1086/376774}, \href
  {https://ui.adsabs.harvard.edu/abs/2003ApJ...594..279G} {594, 279}

\bibitem[\protect\citeauthoryear{{Gordon} et~al.,}{{Gordon}
  et~al.}{2022}]{gordon_flux_cal_2022}
{Gordon} K.~D.,  et~al., 2022, \mn@doi [\aj] {10.3847/1538-3881/ac66dc}, \href
  {https://ui.adsabs.harvard.edu/abs/2022AJ....163..267G} {163, 267}

\bibitem[\protect\citeauthoryear{{Hayden-Pawson} et~al.,}{{Hayden-Pawson}
  et~al.}{2022}]{hayden-pawson_NO_2022}
{Hayden-Pawson} C.,  et~al., 2022, \mn@doi [\mnras] {10.1093/mnras/stac584},
  \href {https://ui.adsabs.harvard.edu/abs/2022MNRAS.512.2867H} {512, 2867}

\bibitem[\protect\citeauthoryear{{Izotov}, {Thuan}  \& {Guseva}}{{Izotov}
  et~al.}{2017}]{izotov_2017}
{Izotov} Y.~I.,  {Thuan} T.~X.,   {Guseva} N.~G.,  2017, \mn@doi [\mnras]
  {10.1093/mnras/stx1629}, \href
  {https://ui.adsabs.harvard.edu/abs/2017MNRAS.471..548I} {471, 548}

\bibitem[\protect\citeauthoryear{{Izotov}, {Thuan}, {Guseva}  \&
  {Liss}}{{Izotov} et~al.}{2018a}]{Izotov2018}
{Izotov} Y.~I.,  {Thuan} T.~X.,  {Guseva} N.~G.,   {Liss} S.~E.,  2018a,
  \mn@doi [\mnras] {10.1093/mnras/stx2478}, \href
  {https://ui.adsabs.harvard.edu/abs/2018MNRAS.473.1956I} {473, 1956}

\bibitem[\protect\citeauthoryear{{Izotov}, {Worseck}, {Schaerer}, {Guseva},
  {Thuan}, {Fricke}  \& {Orlitov{\'a}}}{{Izotov} et~al.}{2018b}]{Izotov2018a}
{Izotov} Y.~I.,  {Worseck} G.,  {Schaerer} D.,  {Guseva} N.~G.,  {Thuan} T.~X.,
   {Fricke} Verhamme A.,   {Orlitov{\'a}} I.,  2018b, \mn@doi [\mnras]
  {10.1093/mnras/sty1378}, \href
  {https://ui.adsabs.harvard.edu/abs/2018MNRAS.478.4851I} {478, 4851}

\bibitem[\protect\citeauthoryear{Izotov, Guseva, Fricke  \& Henkel}{Izotov
  et~al.}{2019}]{izotov_low-redshift_2019}
Izotov Y.~I.,  Guseva N.~G.,  Fricke K.~J.,   Henkel C.,  2019, arXiv e-prints

\bibitem[\protect\citeauthoryear{{Izotov}, {Thuan}  \& {Guseva}}{{Izotov}
  et~al.}{2021a}]{Izotov2021}
{Izotov} Y.~I.,  {Thuan} T.~X.,   {Guseva} N.~G.,  2021a, \mn@doi [\mnras]
  {10.1093/mnras/stab1099}, \href
  {https://ui.adsabs.harvard.edu/abs/2021MNRAS.504.3996I} {504, 3996}

\bibitem[\protect\citeauthoryear{{Izotov}, {Thuan}  \& {Guseva}}{{Izotov}
  et~al.}{2021b}]{izotov_xmp_2021}
{Izotov} Y.~I.,  {Thuan} T.~X.,   {Guseva} N.~G.,  2021b, \mn@doi [\mnras]
  {10.1093/mnras/stab1099}, \href
  {https://ui.adsabs.harvard.edu/abs/2021MNRAS.504.3996I} {504, 3996}

\bibitem[\protect\citeauthoryear{{Jakobsen} et~al.,}{{Jakobsen}
  et~al.}{2022}]{Jakobsen2022}
{Jakobsen} P.,  et~al., 2022, \mn@doi [\aap] {10.1051/0004-6361/202142663},
  \href {https://ui.adsabs.harvard.edu/abs/2022A&A...661A..80J} {661, A80}

\bibitem[\protect\citeauthoryear{{Johnson}, {Leja}, {Conroy}  \&
  {Speagle}}{{Johnson} et~al.}{2021}]{2021ApJS..254...22J}
{Johnson} B.~D.,  {Leja} J.,  {Conroy} C.,   {Speagle} J.~S.,  2021, \mn@doi
  [\apjs] {10.3847/1538-4365/abef67}, \href
  {https://ui.adsabs.harvard.edu/abs/2021ApJS..254...22J} {254, 22}

\bibitem[\protect\citeauthoryear{Kewley, Nicholls  \& Sutherland}{Kewley
  et~al.}{2019}]{kewley_understanding_2019}
Kewley L.~J.,  Nicholls D.~C.,   Sutherland R.~S.,  2019, \mn@doi [Annual
  Review of Astronomy and Astrophysics] {10.1146/annurev-astro-081817-051832},
  57, 511

\bibitem[\protect\citeauthoryear{{Kojima} et~al.,}{{Kojima}
  et~al.}{2020}]{kojima_2020_empress}
{Kojima} T.,  et~al., 2020, \mn@doi [\apj] {10.3847/1538-4357/aba047}, \href
  {https://ui.adsabs.harvard.edu/abs/2020ApJ...898..142K} {898, 142}

\bibitem[\protect\citeauthoryear{{Lamareille}}{{Lamareille}}{2010}]{Lamareille2010}
{Lamareille} F.,  2010, \mn@doi [\aap] {10.1051/0004-6361/200913168}, \href
  {https://ui.adsabs.harvard.edu/abs/2010A&A...509A..53L} {509, A53}

\bibitem[\protect\citeauthoryear{{Langan}, {Ceverino}  \& {Finlator}}{{Langan}
  et~al.}{2020}]{Langan2020}
{Langan} I.,  {Ceverino} D.,   {Finlator} K.,  2020, \mn@doi [\mnras]
  {10.1093/mnras/staa880}, \href
  {https://ui.adsabs.harvard.edu/abs/2020MNRAS.494.1988L} {494, 1988}

\bibitem[\protect\citeauthoryear{Lilly, Carollo, Pipino, Renzini  \&
  Peng}{Lilly et~al.}{2013}]{lilly_gas_2013}
Lilly S.~J.,  Carollo C.~M.,  Pipino A.,  Renzini A.,   Peng Y.,  2013, \mn@doi
  [\apj] {10.1088/0004-637X/772/2/119}, 772, 119

\bibitem[\protect\citeauthoryear{Luridiana, Morisset  \& Shaw}{Luridiana
  et~al.}{2012}]{luridiana_pyneb_2012}
Luridiana V.,  Morisset C.,   Shaw R.~A.,  2012, in {IAU} {Symposium}. pp
  422--423, \mn@doi{10.1017/S1743921312011738}

\bibitem[\protect\citeauthoryear{Luridiana, Morisset  \& Shaw}{Luridiana
  et~al.}{2015}]{luridiana_pyneb_2015}
Luridiana V.,  Morisset C.,   Shaw R.~A.,  2015, \mn@doi [\aap]
  {10.1051/0004-6361/201323152}, 573, A42

\bibitem[\protect\citeauthoryear{Ma, Hopkins, Faucher-Giguère, Zolman,
  Muratov, Kereš  \& Quataert}{Ma et~al.}{2016}]{ma_origin_2016}
Ma X.,  Hopkins P.~F.,  Faucher-Giguère C.-A.,  Zolman N.,  Muratov A.~L.,
  Kereš D.,   Quataert E.,  2016, \mn@doi [\mnras] {10.1093/mnras/stv2659},
  456, 2140

\bibitem[\protect\citeauthoryear{Madau \& Dickinson}{Madau \&
  Dickinson}{2014}]{madau_cosmic_2014}
Madau P.,  Dickinson M.,  2014, \mn@doi [\araa]
  {10.1146/annurev-astro-081811-125615}, 52, 415

\bibitem[\protect\citeauthoryear{{Mahler} et~al.,}{{Mahler}
  et~al.}{2022}]{mahler_lensmodel_smacs_2022}
{Mahler} G.,  et~al., 2022, arXiv e-prints, \href
  {https://ui.adsabs.harvard.edu/abs/2022arXiv220707101M} {p. arXiv:2207.07101}

\bibitem[\protect\citeauthoryear{Maiolino \& Mannucci}{Maiolino \&
  Mannucci}{2019}]{maiolino_re_2019}
Maiolino R.,  Mannucci F.,  2019, \mn@doi [\aapr] {10.1007/s00159-018-0112-2},
  27, 3

\bibitem[\protect\citeauthoryear{Maiolino et~al.,}{Maiolino
  et~al.}{2008}]{maiolino_amaze_2008}
Maiolino R.,  et~al., 2008, \mn@doi [\aap] {10.1051/0004-6361:200809678}, 488,
  463

\bibitem[\protect\citeauthoryear{{Mann} \& {Ebeling}}{{Mann} \&
  {Ebeling}}{2012}]{Mann2012}
{Mann} A.~W.,  {Ebeling} H.,  2012, \mn@doi [\mnras]
  {10.1111/j.1365-2966.2011.20170.x}, \href
  {https://ui.adsabs.harvard.edu/abs/2012MNRAS.420.2120M} {420, 2120}

\bibitem[\protect\citeauthoryear{Mannucci et~al.,}{Mannucci
  et~al.}{2009}]{mannucci_lsd_2009}
Mannucci F.,  et~al., 2009, \mn@doi [\mnras]
  {10.1111/j.1365-2966.2009.15185.x}, 398, 1915

\bibitem[\protect\citeauthoryear{Mannucci, Cresci, Maiolino, Marconi  \&
  Gnerucci}{Mannucci et~al.}{2010}]{mannucci_fundamental_2010}
Mannucci F.,  Cresci G.,  Maiolino R.,  Marconi A.,   Gnerucci A.,  2010,
  \mn@doi [\mnras] {10.1111/j.1365-2966.2010.17291.x}, 408, 2115

\bibitem[\protect\citeauthoryear{{Mannucci}, {Salvaterra}  \&
  {Campisi}}{{Mannucci} et~al.}{2011}]{Mannucci2011}
{Mannucci} F.,  {Salvaterra} R.,   {Campisi} M.~A.,  2011, \mn@doi [\mnras]
  {10.1111/j.1365-2966.2011.18459.x}, \href
  {https://ui.adsabs.harvard.edu/abs/2011MNRAS.414.1263M} {414, 1263}

\bibitem[\protect\citeauthoryear{{Marinacci} et~al.,}{{Marinacci}
  et~al.}{2018}]{Marinacci2018}
{Marinacci} F.,  et~al., 2018, \mn@doi [\mnras] {10.1093/mnras/sty2206}, \href
  {https://ui.adsabs.harvard.edu/abs/2018MNRAS.480.5113M} {480, 5113}

\bibitem[\protect\citeauthoryear{Marino et~al.,}{Marino
  et~al.}{2013}]{marino_o3n2_2013}
Marino R.~A.,  et~al., 2013, \mn@doi [\aap] {10.1051/0004-6361/201321956}, 559,
  A114

\bibitem[\protect\citeauthoryear{{McAlpine} et~al.,}{{McAlpine}
  et~al.}{2016}]{McAlpine2016}
{McAlpine} S.,  et~al., 2016, \mn@doi [Astronomy and Computing]
  {10.1016/j.ascom.2016.02.004}, \href
  {https://ui.adsabs.harvard.edu/abs/2016A&C....15...72M} {15, 72}

\bibitem[\protect\citeauthoryear{{Nagao}, {Murayama}  \& {Taniguchi}}{{Nagao}
  et~al.}{2001}]{Nagao2001}
{Nagao} T.,  {Murayama} T.,   {Taniguchi} Y.,  2001, \mn@doi [\apj]
  {10.1086/319062}, \href
  {https://ui.adsabs.harvard.edu/abs/2001ApJ...549..155N} {549, 155}

\bibitem[\protect\citeauthoryear{Nagao, Maiolino  \& Marconi}{Nagao
  et~al.}{2006}]{nagao_gas_2006}
Nagao T.,  Maiolino R.,   Marconi A.,  2006, \mn@doi [\aap]
  {10.1051/0004-6361:20065216}, 459, 85

\bibitem[\protect\citeauthoryear{{Naiman} et~al.,}{{Naiman}
  et~al.}{2018}]{Naiman2018}
{Naiman} J.~P.,  et~al., 2018, \mn@doi [\mnras] {10.1093/mnras/sty618}, \href
  {https://ui.adsabs.harvard.edu/abs/2018MNRAS.477.1206N} {477, 1206}

\bibitem[\protect\citeauthoryear{{Nakajima} \& {Maiolino}}{{Nakajima} \&
  {Maiolino}}{2022}]{Nakajima_Maiolino22}
{Nakajima} K.,  {Maiolino} R.,  2022, \mn@doi [\mnras]
  {10.1093/mnras/stac1242}, \href
  {https://ui.adsabs.harvard.edu/abs/2022MNRAS.513.5134N} {513, 5134}

\bibitem[\protect\citeauthoryear{{Nakajima} \& {Ouchi}}{{Nakajima} \&
  {Ouchi}}{2014}]{Nakajima14}
{Nakajima} K.,  {Ouchi} M.,  2014, \mn@doi [\mnras] {10.1093/mnras/stu902},
  \href {https://ui.adsabs.harvard.edu/abs/2014MNRAS.442..900N} {442, 900}

\bibitem[\protect\citeauthoryear{{Nakajima}, {Ellis}, {Robertson}, {Tang}  \&
  {Stark}}{{Nakajima} et~al.}{2020}]{Nakajima20}
{Nakajima} K.,  {Ellis} R.~S.,  {Robertson} B.~E.,  {Tang} M.,   {Stark} D.~P.,
   2020, \mn@doi [\apj] {10.3847/1538-4357/ab6604}, \href
  {https://ui.adsabs.harvard.edu/abs/2020ApJ...889..161N} {889, 161}

\bibitem[\protect\citeauthoryear{{Nakajima} et~al.,}{{Nakajima}
  et~al.}{2022}]{nakajima_empress_2022}
{Nakajima} K.,  et~al., 2022, arXiv e-prints, \href
  {https://ui.adsabs.harvard.edu/abs/2022arXiv220602824N} {p. arXiv:2206.02824}

\bibitem[\protect\citeauthoryear{{Nelson} et~al.,}{{Nelson}
  et~al.}{2018}]{Nelson2018}
{Nelson} D.,  et~al., 2018, \mn@doi [\mnras] {10.1093/mnras/stx3040}, \href
  {https://ui.adsabs.harvard.edu/abs/2018MNRAS.475..624N} {475, 624}

\bibitem[\protect\citeauthoryear{{Nelson} et~al.,}{{Nelson}
  et~al.}{2019}]{nelson_TNG_2019}
{Nelson} D.,  et~al., 2019, \mn@doi [Computational Astrophysics and Cosmology]
  {10.1186/s40668-019-0028-x}, \href
  {https://ui.adsabs.harvard.edu/abs/2019ComAC...6....2N} {6, 2}

\bibitem[\protect\citeauthoryear{Nicholls, Dopita, Sutherland, Kewley  \&
  Palay}{Nicholls et~al.}{2013}]{nicholls_measuring_2013}
Nicholls D.~C.,  Dopita M.~A.,  Sutherland R.~S.,  Kewley L.~J.,   Palay E.,
  2013, \mn@doi [\apjs] {10.1088/0067-0049/207/2/21}, 207, 21

\bibitem[\protect\citeauthoryear{Noeske et~al.,}{Noeske
  et~al.}{2007}]{noeske_star_2007}
Noeske K.~G.,  et~al., 2007, \mn@doi [\apjl] {10.1086/517927}, 660, L47

\bibitem[\protect\citeauthoryear{Onodera et~al.,}{Onodera
  et~al.}{2016}]{onodera_ism_2016}
Onodera M.,  et~al., 2016, \mn@doi [\apj] {10.3847/0004-637X/822/1/42}, 822, 42

\bibitem[\protect\citeauthoryear{Osterbrock \& Ferland}{Osterbrock \&
  Ferland}{2006}]{osterbrock_astrophysics_2006}
Osterbrock D.~E.,  Ferland G.~J.,  2006, Astrophysics of {Gaseous} {Nebulae}
  and {Active} {Galactic} {Nuclei}, 2nd edn.
University Science Books

\bibitem[\protect\citeauthoryear{Palay, Nahar, Pradhan  \& Eissner}{Palay
  et~al.}{2012}]{palay_improved_2012}
Palay E.,  Nahar S.~N.,  Pradhan A.~K.,   Eissner W.,  2012, \mn@doi [\mnras]
  {10.1111/j.1745-3933.2012.01252.x}, 423, L35

\bibitem[\protect\citeauthoryear{{Pallottini} et~al.,}{{Pallottini}
  et~al.}{2022}]{pallottini_serra_2022}
{Pallottini} A.,  et~al., 2022, \mn@doi [\mnras] {10.1093/mnras/stac1281},
  \href {https://ui.adsabs.harvard.edu/abs/2022MNRAS.513.5621P} {513, 5621}

\bibitem[\protect\citeauthoryear{{Pascale} et~al.,}{{Pascale}
  et~al.}{2022}]{Pascale2022}
{Pascale} M.,  et~al., 2022, arXiv e-prints, \href
  {https://ui.adsabs.harvard.edu/abs/2022arXiv220707102P} {p. arXiv:2207.07102}

\bibitem[\protect\citeauthoryear{Patrício, Christensen, Rhodin, Cañameras  \&
  Lara-López}{Patrício et~al.}{2018}]{patricio_testing_2018}
Patrício V.,  Christensen L.,  Rhodin H.,  Cañameras R.,   Lara-López M.~A.,
   2018, \mn@doi [\mnras] {10.1093/mnras/sty2508}, 481, 3520

\bibitem[\protect\citeauthoryear{{Perrin}, {Long}, {Sivaramakrishnan},
  {Lajoie}, {Elliot}, {Pueyo}  \& {Albert}}{{Perrin}
  et~al.}{2015}]{2015ascl.soft04007P}
{Perrin} M.~D.,  {Long} J.,  {Sivaramakrishnan} A.,  {Lajoie} C.-P.,  {Elliot}
  E.,  {Pueyo} L.,   {Albert} L.,  2015, {WebbPSF: James Webb Space Telescope
  PSF Simulation Tool}, Astrophysics Source Code Library, record ascl:1504.007
  (\mn@eprint {ascl} {1504.007})

\bibitem[\protect\citeauthoryear{Pettini \& Pagel}{Pettini \&
  Pagel}{2004}]{pettini_oiiinii_2004}
Pettini M.,  Pagel B. E.~J.,  2004, \mn@doi [\mnras]
  {10.1111/j.1365-2966.2004.07591.x}, 348, L59

\bibitem[\protect\citeauthoryear{{Pillepich} et~al.,}{{Pillepich}
  et~al.}{2018}]{Pillepich2018}
{Pillepich} A.,  et~al., 2018, \mn@doi [\mnras] {10.1093/mnras/stx3112}, \href
  {https://ui.adsabs.harvard.edu/abs/2018MNRAS.475..648P} {475, 648}

\bibitem[\protect\citeauthoryear{Pilyugin \& Grebel}{Pilyugin \&
  Grebel}{2016}]{pilyugin_new_2016}
Pilyugin L.~S.,  Grebel E.~K.,  2016, \mn@doi [\mnras] {10.1093/mnras/stw238},
  457, 3678

\bibitem[\protect\citeauthoryear{Pilyugin, Mattsson, Vílchez  \&
  Cedrés}{Pilyugin et~al.}{2009}]{pilyugin_electron_2009}
Pilyugin L.~S.,  Mattsson L.,  Vílchez J.~M.,   Cedrés B.,  2009, \mn@doi
  [\mnras] {10.1111/j.1365-2966.2009.15182.x}, 398, 485

\bibitem[\protect\citeauthoryear{{Pontoppidan} et~al.,}{{Pontoppidan}
  et~al.}{2022}]{pontoppidan_ERO_2022}
{Pontoppidan} K.,  et~al., 2022, arXiv e-prints, \href
  {https://ui.adsabs.harvard.edu/abs/2022arXiv220713067P} {p. arXiv:2207.13067}

\bibitem[\protect\citeauthoryear{{Popesso} et~al.,}{{Popesso}
  et~al.}{2022}]{Popesso2022}
{Popesso} P.,  et~al., 2022, arXiv e-prints, \href
  {https://ui.adsabs.harvard.edu/abs/2022arXiv220310487P} {p. arXiv:2203.10487}

\bibitem[\protect\citeauthoryear{{Repp} \& {Ebeling}}{{Repp} \&
  {Ebeling}}{2018}]{Repp2018}
{Repp} A.,  {Ebeling} H.,  2018, \mn@doi [\mnras] {10.1093/mnras/sty1489},
  \href {https://ui.adsabs.harvard.edu/abs/2018MNRAS.479..844R} {479, 844}

\bibitem[\protect\citeauthoryear{{Repp}, {Ebeling}  \& {Richard}}{{Repp}
  et~al.}{2016}]{Repp2016}
{Repp} A.,  {Ebeling} H.,   {Richard} J.,  2016, \mn@doi [\mnras]
  {10.1093/mnras/stw002}, \href
  {https://ui.adsabs.harvard.edu/abs/2016MNRAS.457.1399R} {457, 1399}

\bibitem[\protect\citeauthoryear{{Riffel} et~al.,}{{Riffel}
  et~al.}{2021}]{Riffel2021}
{Riffel} R.~A.,  et~al., 2021, \mn@doi [\mnras] {10.1093/mnrasl/slaa194}, \href
  {https://ui.adsabs.harvard.edu/abs/2021MNRAS.501L..54R} {501, L54}

\bibitem[\protect\citeauthoryear{Sanders et~al.,}{Sanders
  et~al.}{2016}]{sanders_mosdef_2016}
Sanders R.~L.,  et~al., 2016, \mn@doi [\apj] {10.3847/0004-637X/816/1/23}, 816,
  23

\bibitem[\protect\citeauthoryear{{Sanders} et~al.,}{{Sanders}
  et~al.}{2020}]{2020MNRAS.491.1427S}
{Sanders} R.~L.,  et~al., 2020, \mn@doi [\mnras] {10.1093/mnras/stz3032}, \href
  {https://ui.adsabs.harvard.edu/abs/2020MNRAS.491.1427S} {491, 1427}

\bibitem[\protect\citeauthoryear{{Sanders} et~al.,}{{Sanders}
  et~al.}{2021}]{sanders_mosdef_mzr_2021}
{Sanders} R.~L.,  et~al., 2021, \mn@doi [\apj] {10.3847/1538-4357/abf4c1},
  \href {https://ui.adsabs.harvard.edu/abs/2021ApJ...914...19S} {914, 19}

\bibitem[\protect\citeauthoryear{{Sandles}, {Curtis-Lake}, {Charlot},
  {Chevallard}  \& {Maiolino}}{{Sandles} et~al.}{2022}]{sandles_2022}
{Sandles} L.,  {Curtis-Lake} E.,  {Charlot} S.,  {Chevallard} J.,   {Maiolino}
  R.,  2022, \mn@doi [\mnras] {10.1093/mnras/stac1999}, \href
  {https://ui.adsabs.harvard.edu/abs/2022MNRAS.515.2951S} {515, 2951}

\bibitem[\protect\citeauthoryear{{Schaerer}, {Marques-Chaves}, {Oesch},
  {Naidu}, {Barrufet}, {Izotov}, {Guseva}  \& {Brammer}}{{Schaerer}
  et~al.}{2022}]{Schaerer22}
{Schaerer} D.,  {Marques-Chaves} R.,  {Oesch} P.,  {Naidu} R.,  {Barrufet} L.,
  {Izotov} Y.~I.,  {Guseva} N.~G.,   {Brammer} G.,  2022, arXiv e-prints, \href
  {https://ui.adsabs.harvard.edu/abs/2022arXiv220710034S} {p. arXiv:2207.10034}

\bibitem[\protect\citeauthoryear{{Schaye} et~al.,}{{Schaye}
  et~al.}{2015}]{Schaye_EAGLE_2015}
{Schaye} J.,  et~al., 2015, \mn@doi [\mnras] {10.1093/mnras/stu2058}, \href
  {https://ui.adsabs.harvard.edu/abs/2015MNRAS.446..521S} {446, 521}

\bibitem[\protect\citeauthoryear{{Schneider}, {Hunt}  \&
  {Valiante}}{{Schneider} et~al.}{2016}]{Schneider16}
{Schneider} R.,  {Hunt} L.,   {Valiante} R.,  2016, \mn@doi [\mnras]
  {10.1093/mnras/stw114}, \href
  {https://ui.adsabs.harvard.edu/abs/2016MNRAS.457.1842S} {457, 1842}

\bibitem[\protect\citeauthoryear{Shapley et~al.,}{Shapley
  et~al.}{2015}]{shapley_mosdef_2015}
Shapley A.~E.,  et~al., 2015, \mn@doi [\apj] {10.1088/0004-637X/801/2/88}, 801,
  88

\bibitem[\protect\citeauthoryear{{Shapley} et~al.,}{{Shapley}
  et~al.}{2017}]{2017ApJ...846L..30S}
{Shapley} A.~E.,  et~al., 2017, \mn@doi [\apjl] {10.3847/2041-8213/aa8815},
  \href {https://ui.adsabs.harvard.edu/abs/2017ApJ...846L..30S} {846, L30}

\bibitem[\protect\citeauthoryear{{Shapley} et~al.,}{{Shapley}
  et~al.}{2021}]{2021jwst.prop.1914S}
{Shapley} A.~E.,  et~al., 2021, {The AURORA Survey: First Direct Metallicity
  Calibrations at High Redshift}, JWST Proposal. Cycle 1, ID. \#1914

\bibitem[\protect\citeauthoryear{Shivaei et~al.,}{Shivaei
  et~al.}{2020}]{shivaei_mosdef_2020}
Shivaei I.,  et~al., 2020, \mn@doi [The Astrophysical Journal]
  {10.3847/1538-4357/aba35e}, 899, 117

\bibitem[\protect\citeauthoryear{{Springel} et~al.,}{{Springel}
  et~al.}{2018}]{Springel2018}
{Springel} V.,  et~al., 2018, \mn@doi [\mnras] {10.1093/mnras/stx3304}, \href
  {https://ui.adsabs.harvard.edu/abs/2018MNRAS.475..676S} {475, 676}

\bibitem[\protect\citeauthoryear{Stasińska}{Stasińska}{2002}]{stasinska_abundance_2002}
Stasińska G.,  2002, Cosmochemistry. The melting pot of the elements, pp
  115--170

\bibitem[\protect\citeauthoryear{Steidel et~al.,}{Steidel
  et~al.}{2014}]{steidel_strong_2014}
Steidel C.~C.,  et~al., 2014, \mn@doi [\apj] {10.1088/0004-637X/795/2/165},
  795, 165

\bibitem[\protect\citeauthoryear{Strom, Steidel, Rudie, Trainor, Pettini  \&
  Reddy}{Strom et~al.}{2017}]{strom_nebular_2017}
Strom A.~L.,  Steidel C.~C.,  Rudie G.~C.,  Trainor R.~F.,  Pettini M.,   Reddy
  N.~A.,  2017, \mn@doi [\apj] {10.3847/1538-4357/836/2/164}, 836, 164

\bibitem[\protect\citeauthoryear{{Strom}, {Steidel}, {Rudie}, {Trainor}  \&
  {Pettini}}{{Strom} et~al.}{2018}]{Strom18}
{Strom} A.~L.,  {Steidel} C.~C.,  {Rudie} G.~C.,  {Trainor} R.~F.,   {Pettini}
  M.,  2018, \mn@doi [\apj] {10.3847/1538-4357/aae1a5}, \href
  {https://ui.adsabs.harvard.edu/abs/2018ApJ...868..117S} {868, 117}

\bibitem[\protect\citeauthoryear{{Strom}, {Rudie}, {Chen}, {Law}, {Maseda},
  {Steidel}  \& {Trainor}}{{Strom} et~al.}{2021}]{2021jwst.prop.2593S}
{Strom} A.~L.,  {Rudie} G.~C.,  {Chen} Y.,  {Law} D.~R.,  {Maseda} M.,
  {Steidel} C.~C.,   {Trainor} R.~F.,  2021, {CECILIA: A direct-method
  metallicity calibration for Cosmic Noon through the Epoch of Reionization},
  JWST Proposal. Cycle 1, ID. \#2593

\bibitem[\protect\citeauthoryear{{Tacchella} et~al.,}{{Tacchella}
  et~al.}{2022a}]{2022ApJ...927..170T}
{Tacchella} S.,  et~al., 2022a, \mn@doi [\apj] {10.3847/1538-4357/ac4cad},
  \href {https://ui.adsabs.harvard.edu/abs/2022ApJ...927..170T} {927, 170}

\bibitem[\protect\citeauthoryear{{Tacchella} et~al.,}{{Tacchella}
  et~al.}{2022b}]{tacchella_stellar_pop_2022}
{Tacchella} S.,  et~al., 2022b, \mn@doi [\apj] {10.3847/1538-4357/ac4cad},
  \href {https://ui.adsabs.harvard.edu/abs/2022ApJ...927..170T} {927, 170}

\bibitem[\protect\citeauthoryear{{Tang}, {Stark}, {Chevallard}  \&
  {Charlot}}{{Tang} et~al.}{2019}]{tang_2019}
{Tang} M.,  {Stark} D.~P.,  {Chevallard} J.,   {Charlot} S.,  2019, \mn@doi
  [\mnras] {10.1093/mnras/stz2236}, \href
  {https://ui.adsabs.harvard.edu/abs/2019MNRAS.489.2572T} {489, 2572}

\bibitem[\protect\citeauthoryear{{Topping}, {Shapley}, {Reddy}, {Sanders},
  {Coil}, {Kriek}, {Mobasher}  \& {Siana}}{{Topping}
  et~al.}{2020}]{topping_mosdef-lris_2020_i}
{Topping} M.~W.,  {Shapley} A.~E.,  {Reddy} N.~A.,  {Sanders} R.~L.,  {Coil}
  A.~L.,  {Kriek} M.,  {Mobasher} B.,   {Siana} B.,  2020, \mn@doi [\mnras]
  {10.1093/mnras/staa1410}, \href
  {https://ui.adsabs.harvard.edu/abs/2020MNRAS.495.4430T} {495, 4430}

\bibitem[\protect\citeauthoryear{{Torrey} et~al.,}{{Torrey}
  et~al.}{2019}]{Torrey2019}
{Torrey} P.,  et~al., 2019, \mn@doi [\mnras] {10.1093/mnras/stz243}, \href
  {https://ui.adsabs.harvard.edu/abs/2019MNRAS.484.5587T} {484, 5587}

\bibitem[\protect\citeauthoryear{Tremonti et~al.,}{Tremonti
  et~al.}{2004}]{tremonti_origin_2004}
Tremonti C.~A.,  et~al., 2004, \mn@doi [\apj] {10.1086/423264}, 613, 898

\bibitem[\protect\citeauthoryear{{Troncoso} et~al.,}{{Troncoso}
  et~al.}{2014a}]{Troncoso2014}
{Troncoso} P.,  et~al., 2014a, \mn@doi [\aap] {10.1051/0004-6361/201322099},
  \href {https://ui.adsabs.harvard.edu/abs/2014A&A...563A..58T} {563, A58}

\bibitem[\protect\citeauthoryear{Troncoso et~al.,}{Troncoso
  et~al.}{2014b}]{troncoso_metallicity_2014}
Troncoso P.,  et~al., 2014b, \mn@doi [\aap] {10.1051/0004-6361/201322099}, 563,
  A58

\bibitem[\protect\citeauthoryear{{Witstok}, {Smit}, {Maiolino}, {Curti},
  {Laporte}, {Massey}, {Richard}  \& {Swinbank}}{{Witstok}
  et~al.}{2021}]{2021MNRAS.508.1686W}
{Witstok} J.,  {Smit} R.,  {Maiolino} R.,  {Curti} M.,  {Laporte} N.,  {Massey}
  R.,  {Richard} J.,   {Swinbank} M.,  2021, \mn@doi [\mnras]
  {10.1093/mnras/stab2591}, \href
  {https://ui.adsabs.harvard.edu/abs/2021MNRAS.508.1686W} {508, 1686}

\bibitem[\protect\citeauthoryear{Zahid, Kewley  \& Bresolin}{Zahid
  et~al.}{2011}]{zahid_mass-metallicity_2011}
Zahid H.~J.,  Kewley L.~J.,   Bresolin F.,  2011, \mn@doi [\apj]
  {10.1088/0004-637X/730/2/137}, 730, 137

\bibitem[\protect\citeauthoryear{Zahid, Dima, Kudritzki, Kewley, Geller, Hwang,
  Silverman  \& Kashino}{Zahid et~al.}{2014}]{zahid_universal_2014}
Zahid H.~J.,  Dima G.~I.,  Kudritzki R.-P.,  Kewley L.~J.,  Geller M.~J.,
  Hwang H.~S.,  Silverman J.~D.,   Kashino D.,  2014, \mn@doi [\apj]
  {10.1088/0004-637X/791/2/130}, 791, 130

\makeatother
\end{thebibliography}





\appendix

\section*{Affiliations}
\noindent
{\it
$^{1}$Kavli Institute for Cosmology, University of Cambridge, Madingley Road, Cambridge, CB3 0HA, UK\\
$^{2}$Cavendish Laboratory - Astrophysics Group, University of Cambridge, 19 JJ Thompson Avenue, Cambridge, CB3 0HE, UK\\
$^{3}$Scuola Normale Superiore, Universit\`a di Pisa, Piazza dei Cavalieri 7, I-56126 Pisa, Italy \\
$^{4}$Department of Physics and Astronomy, University College London, Gower Street, London WC1E 6BT, UK\\
$^{5}$Institute of Astronomy, University of Cambridge, Madingley Road, Cambridge, CB3 0HA, UK\\
$^{6}$Sorbonne Universit\'e, UPMC-CNRS, UMR7095, Institut d’Astrophysique de Paris, F-75014, Paris, France\\
$^{7}$National Astronomical Observatory of Japan, 2-21-1 Osawa, Mitaka, Tokyo 181-8588, Japan\\
$^{8}$INAF — Osservatorio Astrofisico di Arcetri, Largo E. Fermi 5, I-50125, Florence, Italy\\
$^{9}$Dipartimento di Fisica e Astronomia, Universit\'a di Firenze, Via G. Sansone 1, I-50125, Sesto Fiorentino (Florence), Italy\\
$^{10}$Centro de Astrobiología (CAB), CSIC–INTA, Cra. de Ajalvir Km. 4, 28850– Torrejón de Ardoz, Madrid, Spain\\
$^{11}$Cosmic Dawn Center, Niels Bohr Institute, University of Copenhagen, Rådmandsgade 62, 2200 Copenhagen N, DK\\
$^{12}$Department of Physics, University of Oxford, Denys Wilkinson Building, Keble Road, Oxford OX1\,3RH, UK\\
$^{13}$Centre for Astrophysics Research, Department of Physics, Astronomy and Mathematics, University of Hertfordshire, Hatfield, AL10 9AB, UK \\
$^{14}$AURA for the European Space Agency, Space Telescope Science Institute, 3700 San Martin Drive, Baltimore, MD 21218, USA\\
$^{15}$Department of Astronomy, University of Wisconsin-Madison, 475 N. Charter St., Madison, WI 53706 USA\\
$^{16}$Astrophysics Research Institute, Liverpool John Moores University, 146 Brownlow Hill, Liverpool L3 5RF, UK\\
}


\bsp	
\label{lastpage}
\end{document}